\documentclass[review]{elsarticle}

\usepackage{amssymb}

\usepackage{booktabs}
\usepackage{graphicx}
\usepackage{adjustbox}
\usepackage{multirow}
\usepackage[table,xcdraw]{xcolor}
\usepackage{url}

\usepackage{graphicx} 
\usepackage{subfigure}
\usepackage{amsmath}
\usepackage[noend]{algpseudocode}
\usepackage{algorithmicx,algorithm}
\usepackage{float}
\usepackage{booktabs}
\usepackage{multirow}
\usepackage{hyperref}
\usepackage{lineno}
\usepackage{amssymb}
\usepackage{xcolor}

\usepackage{xcolor}

\newcommand{\red}[1]{\textcolor{black}{#1}}
\newcommand{\blue}[1]{\textcolor{blue}{#1}}

\newcommand{\rr}[1]{\textcolor{black}{#1}}
\newcommand{\revise}[1]{\textcolor{black}{#1}}
\newcommand{\eg}{\textit{e.g.}}

\newcommand{\ie}{\textit{i.e.}}
\journal{Computer Methods and Programs in Biomedicine}

\begin{document}
\begin{frontmatter}



\title{Learn From Orientation Prior for Radiograph Super-Resolution: Orientation Operator Transformer}


\author[label1,label2]{Yongsong Huang}
\author[label1]{Tomo Miyazaki}
\author[label2]{Xiaofeng Liu}
\author[label3]{Kaiyuan Jiang}
\author[label1]{Zhengmi Tang}
\author[label1]{Shinichiro Omachi}

\affiliation[label1]{organization={Department of Communications Engneering, Graduate School of Engineering, Tohoku University},
            city={Sendai},
            postcode={9808579}, 
            country={Japan}}

\affiliation[label2]{organization={Gordon Center for Medical Imaging, Harvard Medical School},
            city={Boston},
            postcode={02114}, 
            country={USA}}
            
\affiliation[label3]{organization={Department of Surgery, Tohoku University Graduate School of Medicine},
            city={Sendai},
            postcode={80 8575}, 
            country={Japan}}

\begin{abstract}

\textbf{Background and objective:} High-resolution radiographic images play a pivotal role in the early diagnosis and treatment of skeletal muscle-related diseases. It is promising to enhance image quality by introducing single-image super-resolution (SISR) model into the radiology image field. However, the conventional image pipeline, which can learn a mixed mapping between SR and denoising from the color space and inter-pixel patterns, poses a particular challenge for radiographic images with limited pattern features. To address this issue, this paper introduces a novel approach: Orientation Operator Transformer - $O^{2}$former. 
\textbf{Methods:} We incorporate an orientation operator in the encoder to enhance sensitivity to denoising mapping and to integrate orientation prior. Furthermore, we propose a multi-scale feature fusion strategy to amalgamate features captured by different receptive fields with the directional prior, thereby providing a more effective latent representation for the decoder. Based on these innovative components, we propose a transformer-based SISR model, i.e., $O^{2}$former, specifically designed for radiographic images. \textbf{Results:} The experimental results demonstrate that our method achieves the best or second-best performance in the objective metrics compared with the competitors at $\times 4$ upsampling factor. For qualitative, more objective details are observed to be recovered. \textbf{Conclusions:} In this study, we propose a novel framework called $O^{2}$former for radiological image super-resolution tasks, which improves the reconstruction model's performance by introducing an orientation operator and multi-scale feature fusion strategy. Our approach is promising to further promote the radiographic image enhancement field.

\end{abstract}


\end{frontmatter}


\section{Introduction\label{sec.1}}

Radiographs, often referred to as X-rays, are a cornerstone in the realm of musculoskeletal medicine, being instrumental in diagnosing and managing a multitude of diseases. For instance, they play a pivotal role in the evaluation of fractures. They only illustrate the fracture line's location, orientation, and displacement, but also detect joint involvement and identify signs of complications, such as open fractures and infections, e.g., gas gangrene. Radiographs are also adept at revealing indications of osteoporosis, including decreased bone density, trabecular thinning, or cortical thinning\citep{vives2006orthopedic,chen2013age,mc2007vertebral,zhou2015accelerated}. In the context of osteoarthritis, radiographs can pinpoint key features, including joint space narrowing, osteophyte formation, subchondral sclerosis, and cartilage wear\citep{turlington2003radiology,adepu2022biglycan,ying2023inflammation}. Moreover, they can unveil the classic presentations of musculoskeletal tuberculosis, such as bone destruction and periosteal reaction\cite{miyamoto2007pharmacologic,hu2022advance,shen2022dual}. However, the inherent low resolution of radiographic images may lead to diagnostic inaccuracies or oversights. There is a significant need to improve the resolution of these images to ensure more accurate and reliable diagnoses\citep{shin2023multivariable,qiu2022improved,zhu2023feedback,huang2023source}.


To address the issue that the resolution of radiographic images is limited, single image super-resolution (SISR) is attracting increasing attention in this field. The goal of SISR is to recover matched high-resolution (HR) $I_{HR}$ images from degraded low-resolution (LR) $I_{LR}$ images, formalized as follows:

\begin{equation}
I_{LR}=\mathbb{D}\left(I_{H R} ; \delta\right),
\label{eq1}
\end{equation}

\noindent where $\mathbb{D}$ denotes degradation and $\delta$ is the degradation process parameter. SISR is often considered as an  ill-posed problem, considering that we often need to restore paired images from unknowable degradation\cite{wang2020deep,park2003super}. Further, the degradation process is shown in Eq.\ref{eq2}: 

\begin{equation}
\mathbb{D}\left(I_{HR} ; \delta\right)=\left(I_{HR} \otimes \kappa\right) \downarrow_d+n_{\varsigma},
\label{eq2}
\end{equation}
\noindent where $\kappa$ means the blurred kernel in the degradation, and $\downarrow_d$ is the downsampling factor. In general, $n_{\varsigma}$ is considered as additive noise\cite{chen2022real,huang2022infrared}.

With the development of deep learning, it has become popular to use deep learning models in SISR tasks. First, convolutional neural network-based (CNNs-base) approaches were proposed\cite{dong2014learning,kim2016accurate,jiang2021difference}. These methods use CNNs as feature extractors, further learn the nonlinear mapping relationship between LR images and HR images by neural networks, and finally reconstruct the captured features to SR images in latent space. CNN-based models tend to capture more latent representations with the help of attention mechanisms\cite{zhang2018image} and deeper networks\cite{behjati2021overnet}. Then, generative models were introduced to further advance the field. Compared with the CNN-based model, the generative model can recover more image edge detail texture information. The reason behind this is the different optimization paradigms between them. In generative models, the widely studied generative adversarial networks (GANs) are remarkable\cite{goodfellow2020generative,ledig2017photo,huang2021infrared}. The strategy of adversarial generation helps the model to generate diverse samples and no longer relies only on perceptual fields and deep neural networks to improve the generated image quality\cite{wang2018esrgan,wang2021real}. However, the model collapse risk in such methods can bring trouble to the model training\cite{gulrajani2017improved}. Recently, exploding transformer models have become the new paradigm in the field of computer vision. The proposed transformer model is based on the self-attention mechanism, and its excellent ability to capture long-range dependencies enables it to achieve excellent performance in SR tasks\cite{yang2020learning,lu2022transformer}. To balance deep learning-based models between local feature extraction and global information reconstruction, CNNs as encoders combined with transformer decoders are becoming more and more popular\cite{gao2022lightweight}. This kind of approach considers the CNN as an encoder that can have a better perceptive field to capture shallow features, and also takes the excellent long-range information reconstruction capability of the transformer model into consideration, and finally achieves better performance in SISR tasks.

However, these proposed methods can have several challenges in the radiographic field: First, the methods mentioned above focus on the nature image reconstruction, such as RGB-type images. These methods usually focus on the reconstruction between LR-HR pairs more, and the blurred connection between such mappings is not sufficiently explored. However, radiological images usually incorporate more challenging blur in practical applications. For example, breathing undulations are common, making patient displacement relative to the device. This blurring can be widely observed, especially for young or tremor disease patients\citep{qiu2023medical,qiu2022dual,zhu2021residual}. Therefore, the negative influence of noise represented by motion blur on imaging quality is a serious concern. Second, there is a limitation on the feature representation capability of the model by relying on a simple convolutional approach or attention mechanism in the shallow feature extraction. Previous approaches\cite{yang2020learning,lu2022transformer} have tended to improve the decoder, \ie, the transformer model. For CNN-based encoders, improving the ability to capture local information can contribute to the quality of SR images by providing a better latent representation for the decoder.

To further improve the image quality in super-resolution radiography, we propose Orientation Operator Transformer for this task. In summary, our contributions can be summarized as follows:

\begin{itemize}
    \item We first propose Orientation Operator for enhancing CNN-based shallow feature extraction modules. This novel operator focuses on the prior knowledge of horizontal and vertical directions and introduces it into local feature extraction. The different orientation prior helps the encoder to capture shallow features for better latent representation, and  further benefits the decoder in learning better nonlinear mapping for image reconstruction. To the best of our knowledge, it is the first model focusing on the orientation prior in the radiographic super-resolution task.
    \item  We also propose a multi-scale feature fusion strategy for radiographic images. This strategy considers different convolution methods that help to capture shallow features with more diverse local features. The shallow representation with diversity helps the decoder to better reconstruct SR images. Further ablation experiments demonstrate the effectiveness of the feature fusion strategy.
    \item Finally, we propose the end-to-end Orientation Operator Transformer for the super-resolution task in radiographic images. This model includes two components: the CNN encoder for shallow feature extraction and the transformer-based decoder to reconstruct the image by connecting global information. Compared with previous approaches, Orientation Operator Transformer focuses more on non-linear mapping for blurred mapping in radiological image reconstruction and achieves better performance. According to the experimental results, our method achieves better performance compared to competitors' in both objective metrics and qualitative studies.
\end{itemize}

The remainder of this paper is organized as follows: we will present related work on deep learning models and orientation priors in Section.\ref{sec.2}. Section.\ref{sec.3} will present the key components of our proposed Orientation Operator Transformer and detailed information. The qualitative and quantitative evaluation results will be provided in Section.\ref{sec.4}. The conclusions will be included in Section.\ref{sec.5}.

\section{Related Work\label{sec.2}}

In this section, representative work based on deep learning in SISR tasks will be discussed first. These methods will include the following: CNN-based, GAN-based, and transformer-based. Highlighted work will be reviewed and presented. Further, studies in orientation priors will also be shown. Then, the horizontal and vertical prior's application in different visual tasks will be discussed. The details about this prior knowledge in denoising mapping will also be described.

\textbf{Deep learning-based models.} These models aim to reconstruct images by learning a nonlinear mapping between paired data in latent space by using deep learning models. Unlike previous approaches, their advantage is that they do not rely on complicated prior knowledge and can fit the data distribution with the model. The objective function is shown as follows:

\begin{equation}
\hat{\theta}=\underset{\theta}{\arg \min } \mathcal{L}\left(I_{SR}, I_{HR}\right)+\lambda \Phi(\theta),
\end{equation}

where $\mathcal{L}\left(I_{SR}, I_{HR}\right)$ represents the loss function between the generated SR image $I_{SR}$ and the ground truth image $I_{HR},  \Phi(\theta)$ is the regularization term and $\lambda$ is the tradeoff parameter\cite{wang2020deep}. Dong et al.\cite{dong2014learning} introduced the first deep learning model into the SISR domain. The convolution operator used in this shallow model is used for feature extraction in LR images, and finally, the SR images are available by upsampling. This work exceeds other methods at that time and achieves the best performance. Further, Fast Super-Resolution Convolutional Neural Networks (FSRCNN) were proposed by the researchers to minimize the model complexity\cite{dong2015image}. By improving the model, FSRCNN is able to improve computational efficiency while maintaining comparable performance. Such models are more popular in some applications where computing power is limited. The following proposed CNN-based\cite{jiang2021difference} models mainly focus on deepening the layers in the network to improve performance, and introducing attention mechanisms\cite{zhang2018image} is another popular trend.

With the community proposing generative models, such as generative adversarial networks and variational encoders, it has become the trend to introduce these models into SISR tasks\cite{ledig2017photo,huang2021infrared,huang2021het,huang2023target}. The SRGAN model first proposed by Ledig et al. has achieved remarkable performance using an adversarial training strategy. According to the experimental results, the texture details at the edges were well restored for SR images. However, the model collapse risk in this kind of method becomes a challenge in front of researchers. To address this issue, the researchers proposed to use new divergence in the discriminator\cite{gulrajani2017improved,huang2021het}. It was observed from the experimental results that the improved model is more stable during training and also tends to reduce the weakness in mode collapse. Further work focused on the generator model enhancement, and more powerful modules were designed and introduced\cite{huang2022infrared}.

Recent transformer models have attracted the attention in the community from researchers. Such approaches from natural language processing have better decoded latent representations in SISR tasks by their power to capture long-range dependencies\cite{han2020survey,vaswani2017attention}. However, receptive field bounds make these methods fail to achieve the expected performance on local domain extraction. Thus, the CNN-based encoder is used for shallow feature encoding, and the latent output representation is fed to the transformer-based decoder as a new paradigm\cite{gao2022lightweight}.

\textbf{Orientation priors.}  In the lower-level computer vision tasks, the combination of complex elements of the image structure/texture in different orientations (\eg, horizontal and vertical) is considered to benefit the model to fit the data distribution with local features\cite{he2023single,lin2023rethinking}. For example, Nucleus segmentation feeds that prior to the classifier to enhance the response for local features\cite{vo2023mulvernet,dogar2023attention}. Further, in normal image restoration tasks, orientation-aware features are also used as input to fused features to improve the model's representational capability\cite{he2023single}. Moreover, there is also an increasing interest in the denoising field\cite{tsai2022stripformer,sun2015learning,huang2023vicinal} for the orientation feature prior. 

In this study, we aim to introduce the orientation prior to enhance the encoder's ability for local features, which will help the decoder to have better express blur mapping and reconstruct images. Compared to the normal image recovery task, radiological imaging can face more challenges from blur. Previous approaches would also focus on making the model learn to nonlinear mappings, mixing LR-HR and blur. Normal images can learn enough patterns in diverse color patterns to infer the blur orientation between pixels\cite{tsai2022stripformer}, whereas the limited colors and features in radiographic images can introduce more difficulties. To address this issue, further attention to orientation features would be promising to enhance the model's ability to represent blur, which is beneficial to get reconstructed high-quality images. Next, we describe the details for CNN estimation blur from the orientation prior.

Given a blurry image, denoted as $I$, we define the local motion blur kernel at a specific image pixel, represented as $p \in \Omega$ (where $\Omega$ signifies the image region), using a motion vector $\mathbf{m}_{\mathbf{p}}=\left(l_p, o_p\right)$. This vector encapsulates the magnitude and direction of the motion field at point $p$ during the camera shutter's open phase. Each motion vector gives rise to a motion kernel, which possesses non-zero values exclusively along the trajectory of the motion. Consequently, the blurred image can be depicted as $I=k(M) * I_0$, which is to say, the convolution of an underlying sharp image $I_0$ with the non-uniform motion blur kernels $k(M)$, which are dictated by the motion field $M=\left\{\mathbf{m}_{\mathbf{p}}\right\}_{p \in \Omega}$. We also depict the motion vector $\mathbf{m}_{\mathbf{p}}$ as $\left(u_p, v_p\right)$ within the framework of the Cartesian coordinate system, following the transformation:

\begin{equation}
u_p=l_p \cos \left(o_p\right), v_p=l_p \sin \left(o_p\right)
\end{equation}

To summarize, we can use different orientations prior to estimating the blur $\mathbf{m}_{\mathbf{p}}$, considering the two components in the motion blur: $\left(u_p, v_p\right)$. This approach does not make any global parameter assumptions (\eg, homography) for motion, and uses local image regions to predict these kernels \cite{sun2015learning}. In our approach, the encoder's output is the latent representation with better blur estimation, which will be fed to the decoder to learn better nonlinear mappings.

\section{Methodology\label{sec.3}}

In this section, we first provide an overview of the proposed Orientation Operator Transformer: $O^{2}$former, for radiograph super-resolution. Then, we present the detailed configuration of its two main components: the orientation operator and the multi-scale feature fusion strategy. Finally, training strategies will also be demonstrated.

\begin{figure}[t]
\centering
\includegraphics[width=\textwidth]{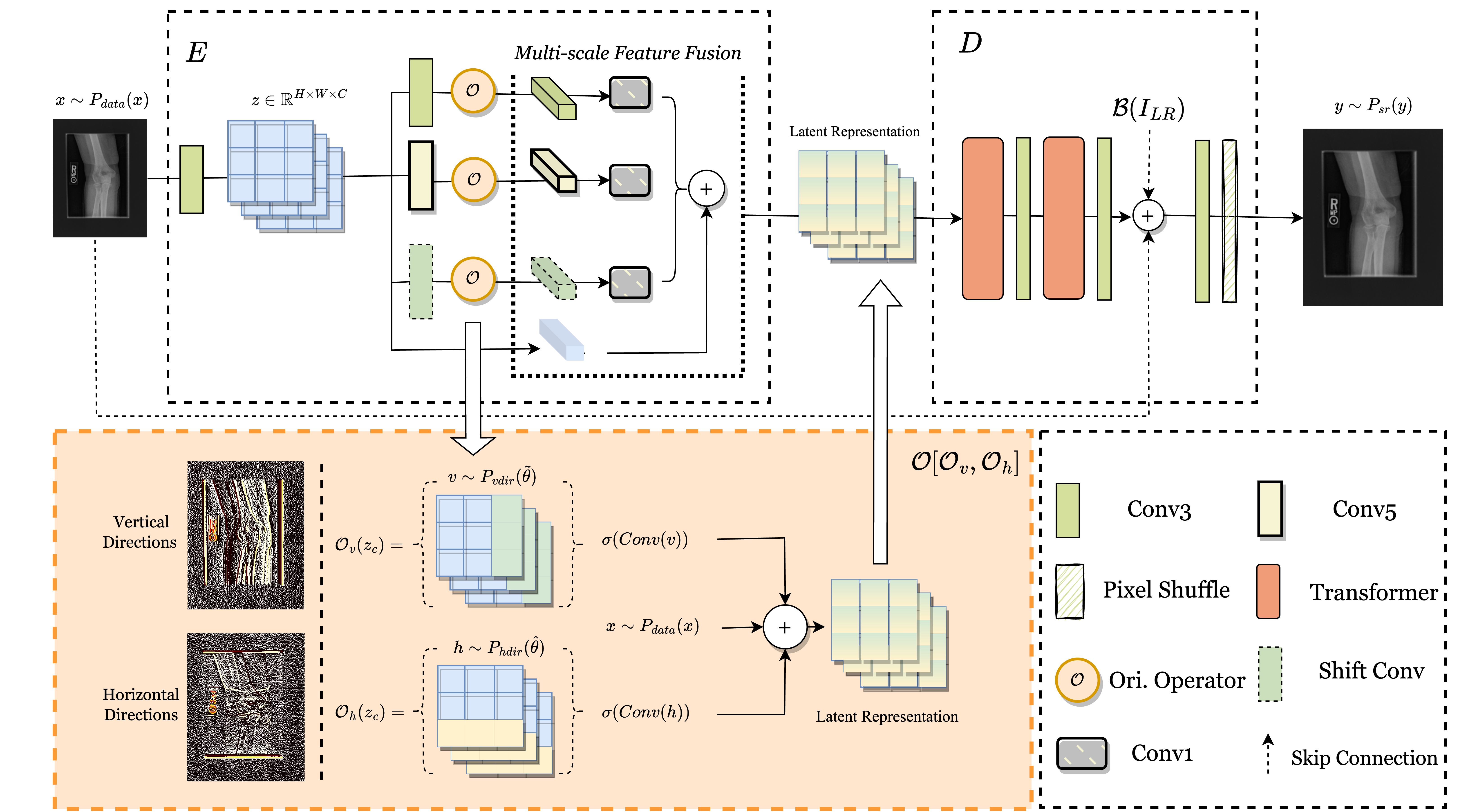}
\caption{\revise{An overview of our proposed $O^{2}$former.} In $O^{2}$former, the encoder $E$ is used to capture shallow local features $z \in \mathbb{R}^{H \times W \times C}$ first, and the $z$ form the output are further fed to the decoder $D$. Our proposed orientation operator $\mathcal{O}$ aims to fuse more orientation priors, both horizontal $h \sim P_{h d i r}(\hat{\theta})$ and vertical $v \sim P_{vdir}(\tilde{\theta})$, for the latent representations. Then, the nonlinear mapping between LR-HR pairs will be learned in the decoder $D$ by parameter optimization. $O^{2}$former with the input $x \sim P_{d a t a}(x)$,  SR image $y \sim P_{s r}(y)$ - output. $\sigma$ denotes the sigmoid function. Zoom in for the best view.}
\label{fig3}
\end{figure}

\subsection{Network Architecture}

To achieve the goal of improving the radiometric SR images quality, we proposed an end-to-end model including encoder $E$ and decoder $D$: $O^{2}$former (see Fig.\ref{fig3}). On the one hand, the local shallow feature is captured and represented as latent representations in the latent space at $E$. In our approach, orientation operator $\mathcal{O}$ and multi-scale feature fusion strategy are proposed, aiming to help the higher quality latent representations to be output. The advantages are as follows: First, different from normal images, the color space and pattern information in radiographic images are poor. Therefore, it is difficult for those models in radiographic images to learn the blurred mapping, which influences the reconstructed image quality, from diverse inter-pixel patterns as normal images. Enhance by $\mathcal{O}$ (more details of the proposed orientation operator is provided in Section.\ref{sec.3.1}), more orientation priors are captured and fused into the shallow feature information. Moreover, the multi-scale feature fusion strategy is introduced to further enhance the model's response to local features at different scales (more details of the proposed multi-scale feature fusion strategy is provided in Section.\ref{sec.3.2}). Finally, $E$ is enhanced by these two novel components to output better latent representations. 

On the other hand, the latent representations from the encoder are fed to the $D$. $O^{2}$former's decoder is proposed based on the transformer model for capturing long-distance dependent pattern information. In $D$, the nonlinear mapping between LR-HR is learned. Previously mixed blurred mappings are also better represented because the input has more blurred estimation priors. As a result, the parameters are optimized for the objective function, and $O^{2}$former will be able to achieve the restoration from LR images to SR images.

\subsection{Orientation Operator\label{sec.3.1}}

The orientation prior is employed as a remarkable image statistical feature widely to enhance model performance in various vision tasks. Considering the complex structural information in the image space domain should include different combinations of textural complexities in various orientations\cite{he2023single,lin2023rethinking}. Previous approaches have further proposed that the image prior knowledge of different orientations, horizontal and vertical, contributes to better fitted data distribution by the model\cite{tsai2022stripformer,sun2015learning}.

In this work, we aim to propose a novel orientation operator $\mathcal{O}\left[\mathcal{O}_v, \mathcal{O}_h\right]$ for radiometric images to capture the image prior that enhances the latent representation quality. The visualization of textures in different directions is shown in Fig.\ref{fig2}, where the blurring priors in shallow features are better represented with the help of this prior knowledge. It will further benefit the decoder to better capture the nonlinear mappings used for reconstruction, such as blurring. To represent the different orientations, $\mathcal{O}$ includes $\mathcal{O}_v$ and $\mathcal{O}_h$, depending on the prior knowledge in vertical and horizontal directions, respectively. For the vertical orientation prior $v \sim P_{vdir}(\tilde{\theta})$, \rr{we first consider the vertically oriented pixel values in the latent variable $\bar{z} \in \mathbb{R}^{H \times W \times C}$ as objects to be extracted ($\bar{z}$ denotes the feature map that has been further extracted by multiple different convolutions, the details will be described in Section.\ref{sec.3.2}.) }, using the pixel-by-pixel operation as follows:

\begin{figure}[t]
\centering
\includegraphics[width=\textwidth]{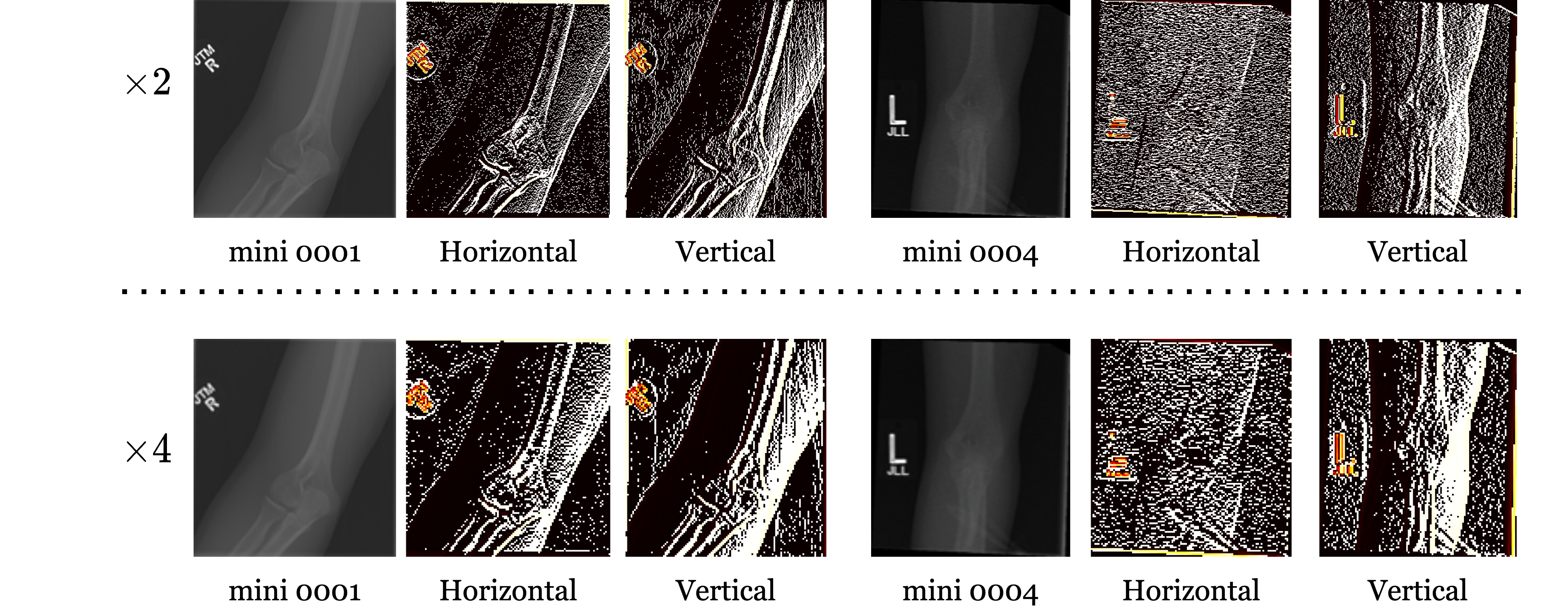}
\caption{Gradients of horizontal and vertical directions from a pair of LR images. The first row is from the LR images in the MURA-mini dataset (the downsampling factor is $\times 2$), and the second row is from the $\times 4$ downsampled data in the same dataset. The streak artifacts in different orientations in the gradient domain. Best viewed in color.}
\label{fig2}
\end{figure}

\begin{equation}
\mathcal{O}_v\left(z_c\right)=\frac{1}{H} \sum_{j \in[0,H)}\left(z_c(W, j)-\Phi^{w} \left(z_c\right)\right)^2
\end{equation}

\rr{First, shallow local features $\bar{z} =[z_1,z_2,z_3...z_c]$ is fed to $\mathcal{O}_v$ and average pooling $\Phi^{w}$ is performed. $c$ denotes the number of channels. $W$, $H$ are width and height, respectively.} Further, we want to minimize the variance between pixels, and the pixel-by-pixel in the feature map $z_c$ is given a variance pooling operation. Our aim is to balance the computational complexity and space direction prior. Similarly, the direction operator $\mathcal{O}_h$ for the horizontal direction is defined as:

\begin{equation}
\mathcal{O}_h\left(z_c\right)=\frac{1}{W} \sum_{i \in[0, W)}\left(z_c(i, H)-\Phi^{h}\left(z_c\right)\right)^2
\end{equation}

\rr{For the horizontal direction, $z_c$ as the input via horizontal averaging pooling $\Phi^{h}$ will produce the initial statistical features for the independent channels. The higher order directional prior will be available by variance pooling.} When directional priors for different directions are captured, we will fuse these priors, $v \sim P_{vdir}(\tilde{\theta})$ and $h \sim P_{h d i r}(\hat{\theta})$, with the latent representations. The fusion details will be described in the multi-scale fusion feature strategy.

\subsection{Multi-scale Feature Fusion Strategy\label{sec.3.2}}

In this section, we describe the multiscale feature fusion strategy $\mathcal{F} $ used in $O^2$ former.
Numerous prior studies\cite{zhang2018residual,li2018multi,tong2017image} have underscored the critical role of hierarchical features from distinct convolutional stages in augmenting SISR. Our multi-scale feature fusion strategy is designed with a hierarchical structure of different convolution modules to generate orientation-aware features across a variety of scales. Unlike previous strategies, our approach focuses on both different convolutional perceptual features and fuses directional priors. A strategic approach is adopted where orientation priors are tasked with holding a rich set of denoising features, while these patterns benefit the decoder in learning the blurring mapping. \revise{Our methodology employs convolution operators of varying scales (\eg, $5 \times 
 5$ and $3 \times 3$) to extract shallow features from images at multiple scales. Following this, the Orientation Operator delineates the orientation prior across these scales, detailed in Section.\ref{sec.3.1}. Subsequent to this, $1 \times 1$ convolution is applied to reshape the feature maps to identical dimensions, allowing for their summation with the shallow feature maps, thereby integrating the multi-scale orientation prior with the shallow features (see Fig.\ref{fig3}). This fusion approach emphasizes the assimilation of multi-scale image orientation priors over the shallow features, advancing beyond previous strategies that targeted solely pixel-level feature disparities.} 

After shallow feature extraction $z \in \mathbb{R}^{H \times W \times C}$, we seek to capture high-level features by employing multiple convolution methods in order to help the model represent more orientation-aware features at different scales. \rr{Specifically, different convolutional approaches are used, including $\text{Conv}_{3 \times 3}$, $\text{Conv}_{5 \times 5}$, and ShiftConv\cite{zhang2022efficient}. The formalization is as follows:}


\begin{equation}
\bar{z}=\left[C_{3 \times 3}(z), C_{5 \times 5}(z), C_{\text {shift }}(z)\right] \odot \mathcal{O}
\label{eq7}
\end{equation}

$C_{3 \times 3}$ and $C_{5 \times 5}$ denote convolution operations utilizing $3 \times 3$ and $5 \times 5$ kernels, respectively. The symbol $\odot$ signifies the linear operation between the convolution kernel and the latent representation. As illustrated in the proposed pipeline, the initial step involves the extraction of shallow features $z$ through multi-scale representations $\left(C_{3 \times 3}, C_{5 \times 5}, C_{s h i f t}\right)$. Subsequently, latent representations at varying scales are input into $\mathcal{O}$, facilitating the capture of perceptual priors across different orientations. These advanced radiographic priors are then integrated with the initial image features $x$ in the next stage, yielding enhanced latent representations. \rr{The formalization is shown below: }

\begin{equation}
\mathcal{F}=\sum\left(\sigma(\bar{z} \odot C_{1 \times 1} ), x\right)
\end{equation}

Particularly, considering the aggregated feature maps generated by Eq.\ref{eq7}, $1\times 1$ convolutional transformations are employed. These transformations are applied independently to feature maps in $\bar{z}$. $\sigma$ represents the sigmoid function. Following the application of the nonlinear mapping function $\sigma$, the feature maps, now enriched with priors and encompassing various scales, undergo normalization. These normalized values are then added to $x$ on a pixel-by-pixel basis, resulting in a latent representation for the decoder $D$ that incorporates priors from multiple orientations. These enhanced representations facilitate the decoder $D$ in more effective learning of the mapping between LR-HR pairs, as well as the denoising mapping.

\subsection{Training Strategy \label{sec.3.3}}

During the training process, the optimization of $O^2$ former is achieved through the application of the $\mathcal{L}_{1}$ loss function and MSE loss $\mathcal{L}_{mse}$. Given a training dataset, denoted as $\left\{I_{LR}^i, I_{H R}^i\right\}_{i=1}^N$, we aim to solve the following optimization problem:

\begin{equation}
\hat{\theta}=\arg \min _\theta\left(\alpha\left[\frac{1}{N} \sum_{i=1}^N\left\|F_\theta\left(I_{L R}^i\right)-I_{H R}^i\right\|_1\right]+\beta \mathcal{L}_{m s e}\right)
\end{equation}

In this formulation, $\theta$ stands for the set of parameters that define our proposed $O^2$ former. The function $F_{\theta}\left(I_{L R}\right)=I_{S R}$ represents the process of reconstructing the SR image from the LR input. Lastly, $N$ is indicative of the total count of images present in the training dataset. $\alpha$ and $\beta$ denote the weights. This optimization strategy is designed to minimize the difference between the reconstructed SR image and the original HR image, thereby enhancing the performance of the $O^{2}$former model.

\section{Experiments\label{sec.4}}

\begin{figure}
\centering
\includegraphics[width=\textwidth]{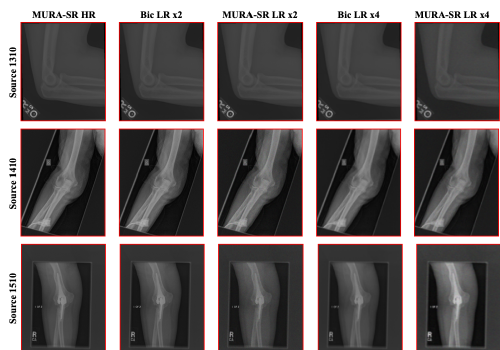}
\caption{Examples of training samples and their corresponding LR images.} 
\label{fig.4}  
\end{figure}

\textbf{Training Dataset:}\label{sec3.1}
In our study, we utilize a widely recognized dataset, MURA-SR~\cite{huang2022rethinking}, to generate training pairs. MURA-SR comprises 4,000 musculoskeletal radiographs of the upper extremity. To emulate real-world scenarios with fewer samples, we curate a subset of 500 images from the MURA-SR dataset to form our training set. Representative samples from this training dataset can be viewed in Fig.\ref{fig.4}, which showcases samples with varying degrees of degradation and downsampling factors. Upon examination of these samples, it becomes evident that complex degradation can result in increased blurring of the edges in LR images, as exemplified by the alphabets in Source 1310. Furthermore, statistical noise can introduce undesirable artifacts into the bone structures, as can be observed in Source 1410.

\textbf{Test Dataset:} The evaluation of these experiments will encompass both objective comparisons and subjective visual assessments. Further details about the datasets are provided below. For synthetic datasets, we utilize two test datasets: MURA-mini and MURA-plus. Both datasets consist of 100 HR images, but differ in terms of the degradation levels of their respective LR images.

\textbf{Training details:} Our model is trained using a batch size of 32, executed on two TITAN X (Pascal) GPUs. The size of the HR patch used during training is set to 96. For optimization, we utilize the Adam optimizer~\cite{kingma2014adam}, with a learning rate set at $1e- 5$. The number of training epochs is established at 1000. \red{$\alpha$ and $\beta$ were set to 1 and 0.1, respectively.} 

\textbf{Evaluation Metrics:} In alignment with the baseline experimental results, our experiments employ widely accepted full reference evaluation metrics: the Peak Signal-to-Noise Ratio (PSNR) and the Structural Similarity Index (SSIM)\cite{wang2004image}. Concurrently, objective metrics are also assessed using non-reference metrics: the Perceptual Index (PI)\cite{blau20182018} and the Naturalness Image Quality Evaluator (NIQE)\cite{mittal2012making}. Quantitative results are evaluated specifically on the luminance channel (Y).

\begin{table}[htbp]
\setlength{\abovecaptionskip}{0cm}
\setlength{\belowcaptionskip}{0.2cm}
    \centering
    \begin{minipage}[t]{0.41\columnwidth}
        \centering
        \caption{Ablation studies of the encoder. Average PSNR $\uparrow$ on MURA-mini datasets with scale factor ×4 are shown.}
        \resizebox{\columnwidth}{!}{%
                \begin{tabular}{@{}c|c|cl@{}}
                \toprule
                Scale                       & Encoder      & \multicolumn{2}{c}{PSNR/dB $\uparrow$} \\ \midrule
                \multirow{4}{*}{$\times 4$} & CNN (w/o)    & 29.66   & \textcolor[RGB]{217,205,144}{Baseline}          \\
                                            & CNN(w)       & 30.06   &                   \\
                                            & Attention & 29.91   &                   \\
                                            & Ours         & 30.24   & \textcolor[RGB]{237,120,74}{0.58 dB $\uparrow$}   \\ \bottomrule
                \end{tabular}%
                }
        \label{tab:sub1}
    \end{minipage}
    \hfill
    \begin{minipage}[t]{0.51\columnwidth}
        \centering
        \caption{Ablation experiments of shallow feature fusion methods. Results for different fusion methods in the MURA-mini dataset $\times 4$ are presented.}
        \resizebox{\columnwidth}{!}{%
        \begin{tabular}{@{}c|c|cc@{}}
        \toprule
        Scale                       & Feature Fusion             & \multicolumn{2}{c}{PSNR/dB $\uparrow$} \\ \midrule
        \multirow{4}{*}{$\times 4$} & concat all                 & 30.14  & \textcolor[RGB]{217,205,144}{Baseline}           \\
                                    & concat all $\& $ skip      & 30.07  &                    \\
                                    & concat (conv3 $\& $ conv5) & 30.09  &                    \\
                                    & sum                     & 30.24  & \textcolor[RGB]{237,120,74}{0.10 dB $\uparrow$}  \\ \bottomrule
        \end{tabular}%
        }
        \label{tab:sub2}
    \end{minipage}
    \label{tab:main}
\end{table}

\subsection{Ablation Studies\label{sec.4.3}}

To delve deeper into the performance of the proposed methods, we conduct an ablation study to analyze their impact on model training. Initially, we illustrate the effectiveness of the proposed $O^{2}$former and encoder. Subsequently, we carry out an ablation experiment to examine the influence of the key components of our architectural design: fusion strategies and multi-scale convolution.

\textbf{Influence of $O^{2}$former \& encoder:} We investigated the performance of various types of shallow feature encoders used in the SR task for radiographic images, as detailed in Tab.\ref{tab:sub1}. Initially, we excluded CNN-based encoders from the model used in this task, resulting in a PSNR score of 29.66 dB. As a baseline model, the experimental results demonstrated that an SR model using only transformers performed sub-optimally. As previously discussed, this is often attributed to the fact that transformer models excel at capturing long-distance dependent information, such as pair data mapping. The limited receptive field presented by smaller patch data restricts performance, which could be enhanced by employing a CNN-based encoder with a larger receptive field.

Our experiments with the CNN-based encoder revealed that the enhanced model outperformed the baseline, improving objective metrics by up to 0.4 dB. This suggests that the CNN-based encoder is more effective in capturing shallow-level features. We then proposed attention-based CNN improvement methods\cite{behjati2023single} and evaluated the performance of the attention mechanism on this baseline approach. The experiments indicated that attention mechanisms, while effective for normal images, face challenges when applied to radiographic images, with a PSNR evaluation metric of 29.91 dB. These methods primarily focus on mixing pattern features between pixels in the latent space. While the model used for normal images can effectively represent LR-HR pairwise mapping and blur mapping due to the diverse color space and pixel-to-pixel pattern information, this information is more limited in radiographic images, presenting challenges for these attention mechanisms.

Finally, our method achieved the best performance in the ablation experiments, with a PSNR score of 30.24 dB. This represents an improvement of 0.58 dB compared to the baseline model, demonstrating the effectiveness of the $O^{2}$former we incorporated into the encoder for this task.

\textbf{Influence of feature fusion strategy:} In this section, our objective is to investigate the approach of fusing features and the selection of the convolution method. As depicted in Tab.\ref{tab:sub2}, we initially attempt to concat feature maps from different convolution methods. This results in a PSNR score of 30.14 dB, indicating an improvement over the baseline that does not rely on feature fusion (refer to Tab.\ref{tab:sub1}). The experiments demonstrate that multi-scale feature extraction is advantageous for enhancing model performance. Additionally, we explored the combination of concat with skip connections, which did not yield competitive performance. We also experimented with selectively choosing parts of the convolutional feature maps for concatenation. The results indicate that omitting shallow features from other perceptual fields is detrimental to performance, with a PSNR score of 30.09 dB.

Finally, we intuitively summed the feature maps on a pixel-by-pixel basis to obtain the latent representations. This operation not only fuses multi-scale feature information but also better preserves the spatial semantic information in the orientations between neighboring pixels a priori. As the experimental results demonstrate, the PSNR is improved to 30.24 dB.

\begin{table}[t]
\setlength{\abovecaptionskip}{0cm}
\setlength{\belowcaptionskip}{0.2cm}
\centering
\caption{The results of adding different convolution methods in $O^{2}$former. Average PSNR$\uparrow$ on MURA-mini datasets with scale factor $\times 4$ are shown.}
\label{tab3}
\resizebox{0.7\columnwidth}{!}{%
\begin{tabular}{@{}c|cccc|cl@{}}
\toprule
Scale                       & Conv. $3 \times 3$ & Conv. $5 \times 5$ & Conv. shift & Skip    & \multicolumn{2}{c}{PSNR/dB $\uparrow$} \\ \midrule
\multirow{4}{*}{$\times 4$} & $\surd$          &                  &               &         & 29.83  & \textcolor[RGB]{217,205,144}{Baseline}          \\
                            & $\surd$          & $\surd$          &               &         & 29.99  &                    \\
                            & $\surd$          & $\surd$          & $\surd$       &         & 30.04  &                    \\
                            & $\surd$          & $\surd$          & $\surd$       & $\surd$ & 30.24  & \textcolor[RGB]{237,120,74}{0.41 dB $\uparrow$} \\ \bottomrule
\end{tabular}%
}
\end{table}

\textbf{Influence of convolution methods:} As outlined in Tab.\ref{tab3}, we built the encoder using convolutional strategies with different perceptual fields. We started with a convolution kernel of size 3, then moved to larger kernels to capture more diverse scale features, improving the PSNR score to 29.99 dB. Incorporating advanced convolution methods\cite{zhang2022efficient} further enhanced the model's ability to represent shallow features, achieving a score of 30.04 dB. We also introduced a skip connection between the shallow features and the final upsampling stage to transfer more information from the shallow patterns. 

The experiments showed that using diverse convolutional approaches allowed the encoder's latent representation to include multi-scale features. The final PSNR score reached 30.24 dB, an improvement of 0.41 dB over the benchmark experiment.

\subsection{Results\label{sec.4.4}}

\begin{table}[t]
\setlength{\abovecaptionskip}{0cm}
\setlength{\belowcaptionskip}{0.2cm}
\centering
\caption{PSNR$\uparrow $ and SSIM$\uparrow $ results of different methods on MURA-mini (mini) \& MURA-plus (plus)  with scale factors of  4 \& 2. Non-reference metrics: NIQE$\downarrow$ and PI$\downarrow$ were also used to evaluate image quality, respectively. There are two test datasets with the same HR images but different degraded LR images (which are more damaged). The \textcolor{red}{red} and \textcolor{blue}{blue} indicate the best and the second-best performance, respectively.} 
\label{tab2}
\resizebox{0.9\textwidth}{!}{%
\begin{tabular}{@{}ccccccccccc@{}}
\toprule
                                                  & \multicolumn{2}{c}{}                                                    & Ours                & SAFMN                           & 
                                                  Shu                     & 
                                                  OverNet                       & FSRCNN                        & RCAN                         & SRCNN   & Bic                           \\ \midrule
\multicolumn{1}{c|}{}                             & \multicolumn{1}{c|}{}                       & \multicolumn{1}{c|}{PSNR} & {\color[HTML]{FE0000} 30.24} &\blue{30.23} & 30.09  & 30.02                         & 29.98                         & 29.70                        & 28.86   & 26.48                         \\
\multicolumn{1}{c|}{}                             & \multicolumn{1}{c|}{}                       & \multicolumn{1}{c|}{SSIM} & 0.9484     & 0.9504                  & {\color[HTML]{3166FF} 0.9506} & 0.9503                        & {\color[HTML]{FE0000} 0.9508} & 0.9493                       & 0.9426  & 0.9362                        \\ \cmidrule(lr){3-3}
\multicolumn{1}{c|}{}                             & \multicolumn{1}{c|}{}                       & \multicolumn{1}{c|}{NIQE} & {\color[HTML]{FE0000} 8.3320} & 9.0689 & {\color[HTML]{3166FF} 8.6955} & 8.9148                        & 9.0839                        & 9.0024                       & 9.5310  & 8.9988                        \\
\multicolumn{1}{c|}{}                             & \multicolumn{1}{c|}{\multirow{-4}{*}{mini}} & \multicolumn{1}{c|}{PI}   & {\color[HTML]{FE0000} 7.6738} &- & {\color[HTML]{3166FF} 7.9935} & 8.0562                        & 8.2511                        & 8.1178                       & 8.6483  & 8.1202                        \\ \cmidrule(l){2-11} 
\multicolumn{1}{c|}{}                             & \multicolumn{1}{c|}{}                       & \multicolumn{1}{c|}{PSNR} & \blue{30.55} &\red{30.57} & {\color[HTML]{3166FF} 30.35}  & 30.22                         & 30.22                         & 29.89                        & 28.99   & 26.80                         \\
\multicolumn{1}{c|}{}                             & \multicolumn{1}{c|}{}                       & \multicolumn{1}{c|}{SSIM} & 0.9417    & 0.9433                   & {\color[HTML]{3166FF} 0.9434} & 0.9432                        & {\color[HTML]{FE0000} 0.9437} & 0.9422                       & 0.9375  & 0.9314                        \\ \cmidrule(lr){3-3}
\multicolumn{1}{c|}{}                             & \multicolumn{1}{c|}{}                       & \multicolumn{1}{c|}{NIQE} & {\color[HTML]{3166FF} 8.7672} &9.5649  & {\color[HTML]{FE0000} 8.4588} & 9.3380                        & 9.5755                        & 9.4267                       & 10.1283 & 9.2759                        \\
\multicolumn{1}{c|}{\multirow{-8}{*}{$\times 4$}} & \multicolumn{1}{c|}{\multirow{-4}{*}{plus}} & \multicolumn{1}{c|}{PI}   & {\color[HTML]{FE0000} 8.0226} &- & 9.4680                        & 8.3886                        & 8.5966                        & 8.4340                       & 9.0154  & {\color[HTML]{3166FF} 8.3265} \\ \midrule
\multicolumn{1}{c|}{}                             & \multicolumn{1}{c|}{}                       & \multicolumn{1}{c|}{PSNR} & {\color[HTML]{FE0000} 31.35} &31.05 & 31.27                         & 29.87                         & 31.30                         & {\color[HTML]{3166FF} 31.32} & 30.17   & 27.60                         \\
\multicolumn{1}{c|}{}                             & \multicolumn{1}{c|}{}                       & \multicolumn{1}{c|}{SSIM} & 0.9372  & 0.9414                     & {\color[HTML]{FE0000} 0.9442} & {\color[HTML]{3166FF} 0.9439} & 0.9434                        & 0.9425                       & 0.9433  & 0.9405                        \\ \cmidrule(lr){3-3}
\multicolumn{1}{c|}{}                             & \multicolumn{1}{c|}{}                       & \multicolumn{1}{c|}{NIQE} & 7.8368  & 8.5484                     & 8.3353                        & {\color[HTML]{3166FF} 7.7891} & 8.1240                        & 8.0038                       & 7.8758  & {\color[HTML]{FE0000} 7.6566} \\
\multicolumn{1}{c|}{}                             & \multicolumn{1}{c|}{\multirow{-4}{*}{mini}} & \multicolumn{1}{c|}{PI}   & {\color[HTML]{3166FF} 7.1446} &- & 7.4418                        & 7.3260                        & 7.3324                        & 7.2628                       & 7.2917  & {\color[HTML]{FE0000} 6.9207} \\ \cmidrule(l){2-11} 
\multicolumn{1}{c|}{}                             & \multicolumn{1}{c|}{}                       & \multicolumn{1}{c|}{PSNR} & 31.85    & 31.72                    & 31.84                         & 30.11                         & {\color[HTML]{3166FF} 31.97}  & {\color[HTML]{FE0000} 32.01} & 30.48   & 27.66                         \\
\multicolumn{1}{c|}{}                             & \multicolumn{1}{c|}{}                       & \multicolumn{1}{c|}{SSIM} & 0.9266   & 0.9316                    & {\color[HTML]{3166FF} 0.9356} & {\color[HTML]{FE0000} 0.9358} & 0.9338                        & 0.9329                       & 0.9347  & 0.9324                        \\ \cmidrule(lr){3-3}
\multicolumn{1}{c|}{}                             & \multicolumn{1}{c|}{}                       & \multicolumn{1}{c|}{NIQE} & 8.2203   & 8.7294                    & 8.1982                        & {\color[HTML]{3166FF} 8.1563} & 8.4326                        & 8.2614                       & 8.1968  & {\color[HTML]{FE0000} 7.8381} \\
\multicolumn{1}{c|}{\multirow{-8}{*}{$\times 2$}} & \multicolumn{1}{c|}{\multirow{-4}{*}{plus}} & \multicolumn{1}{c|}{PI}   & {\color[HTML]{3166FF} 7.3677} &- & 7.4858                        & 7.5938                        & 7.5277                        & 7.4645                       & 7.5491  & {\color[HTML]{FE0000} 7.1892} \\ \bottomrule
\end{tabular}}%
\end{table}

\begin{figure}[H]
	\centering
	\begin{minipage}{\textwidth}		\centerline{\includegraphics[width=\textwidth]{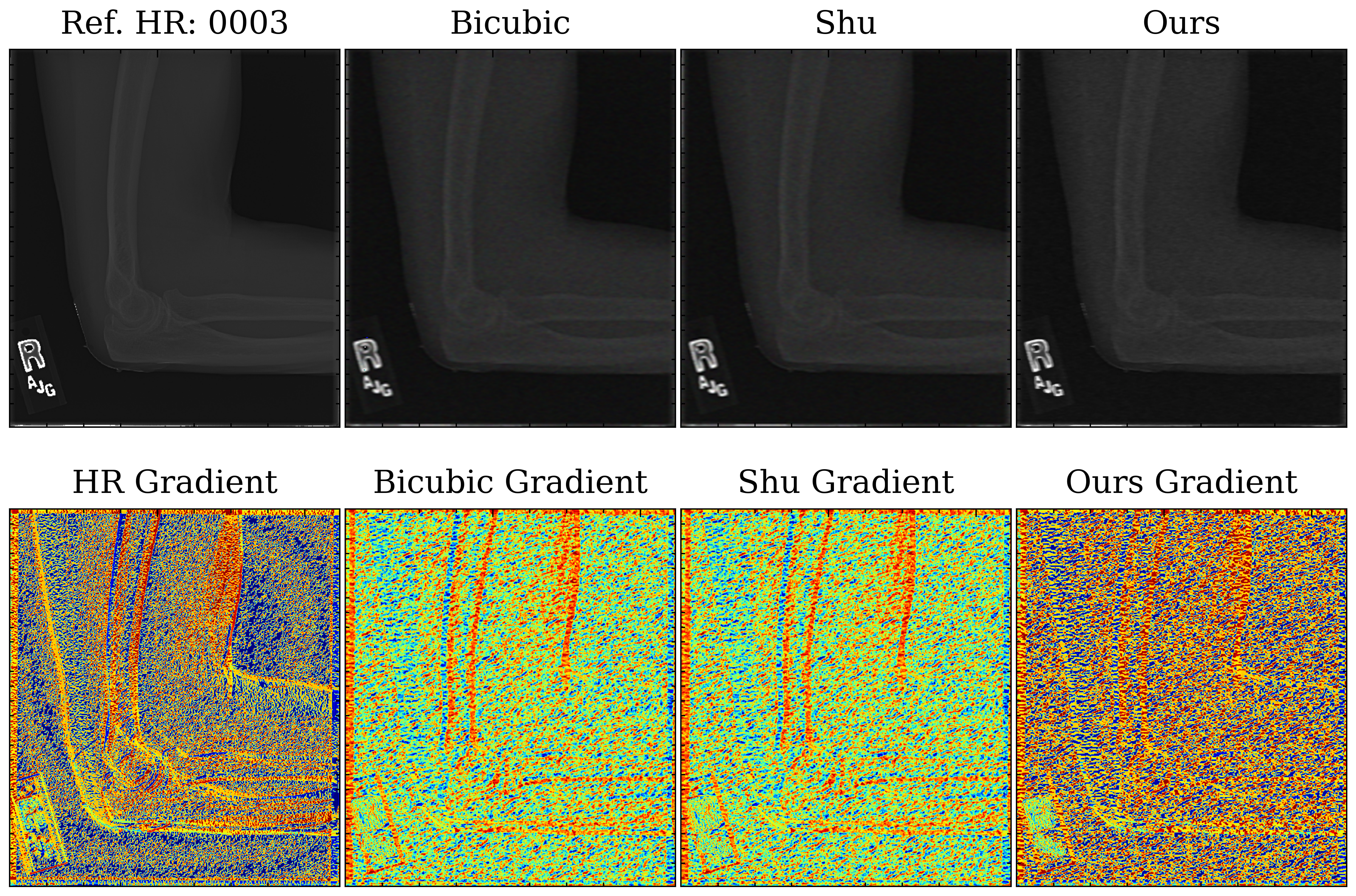}}		\centerline{\includegraphics[width=\textwidth]{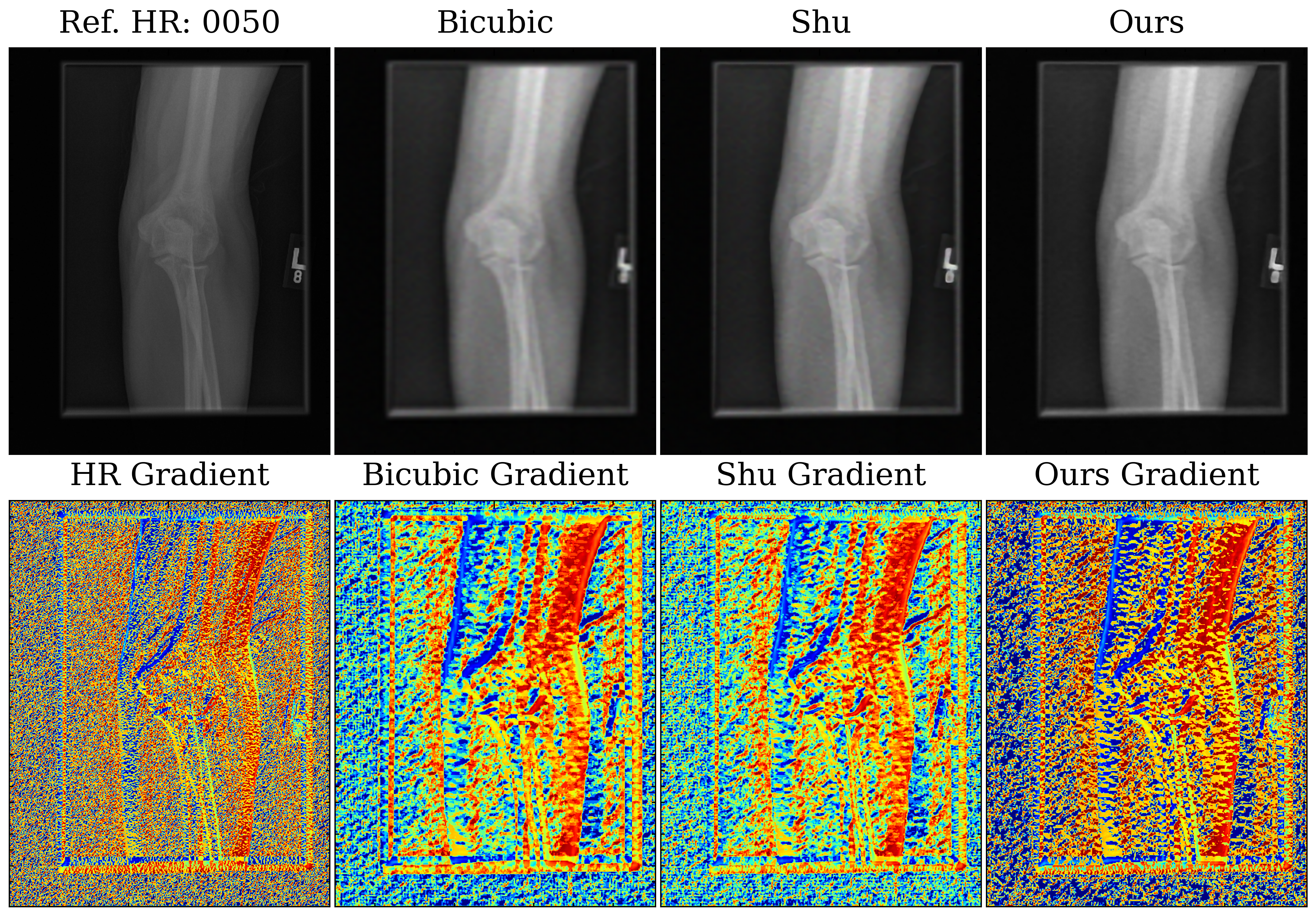}}
	\end{minipage}
	\caption{Gradient visual comparisons of our model and others (Bic \& Shu). The streak artifacts can be easily captured in the gradient domain. Sample data are from the $\times 2$ MURA-mini dataset, and the colorful pattern in the heat map indicates the diversity of detailed textures.}
	\label{fig4}
\end{figure}

\textbf{Quantitative results:} Our method consistently outperforms the state-of-the-art: Shu\cite{sun2022shufflemixer}, SAFMN\cite{sun2023spatially} and other methods (OverNet\cite{behjati2021overnet}, FSRCNN\cite{dong2016accelerating}, RCAN\cite{zhang2018image}, SRCNN\cite{dong2014learning}, Bic) across both MURA-mini and MURA-plus datasets in terms of PSNR, a key indicator of image fidelity. Specifically, for $\times 4$ scaling factor on the MURA-mini dataset, our method achieves a PSNR of 30.24, surpassing Shu's 30.09. This trend continues on the MURA-plus dataset, where our method scores 30.55 in PSNR, again outperforming Shu's 30.35. Furthermore, our method excels in the NIQE and PI metrics, which measure image naturalness and perceptual quality respectively. On the MURA-mini dataset with $\times 4$ scaling factor, our method achieves the best NIQE and PI scores of 8.3320 and 7.6738, respectively, indicating superior performance in terms of image naturalness and perceptual quality. When the scaling factor is reduced to $\times 2$, our method continues to perform well on the MURA-mini dataset, achieving the highest PSNR score of 31.35. This demonstrates the robustness of our method across different scaling factors. However, it's worth noting that the Bic method achieves better performance in certain metrics. This is particularly evident in the PI on the MURA-mini dataset with $\times 2$ scaling factor, where Bic scores the best with 6.9207. These metrics were originally designed for normal images, and the fact that Bic performs well on these metrics suggests it may be preserving certain aspects of the image that are particularly valued in normal images. To further investigate these differences, we plan to conduct a qualitative analysis, visualizing the detail and texture information in the images produced by these methods (see Fig.\ref{fig4}). This will provide a more comprehensive understanding of our method compared to others.


\textbf{Qualitative results:} In this section, we will present the visualization experimental results for more qualitative experiments. First, to address the concerns in the quantitative results for the reconstruction results from the Bic method, we use the Histogram of Oriented Gradient (HOG) operator\cite{dalal2005histograms} to visualize the detailed texture information in the gradient domain. Gradient operators are used following previous work in \cite{huang2021gan}. As shown in Fig.\ref{fig4}, our method would provide more heat map response compared to Bic and Shu. In the $\times 2$ upsampling factor comparison, although the reconstructed image by the Bic method achieves higher scores, the $O^{s}$former's detail recovery is more competitive and reliable in the human subjective visual evaluation. \rr{In comparison to existing other methodologies, our proposed technique elevates the quality of visualization by markedly optimizing the resolution and discernibility between bone structures (see Fig.\ref{fig4}, Ref. HR: 0003) and the contiguous muscular soft tissue (see Fig.\ref{fig4}, Ref. HR: 0050). When benchmarked against established models such as Bic and Shu, it demonstrates comparable, if not superior, performance in terms of discernibility, whilst preserving a gradient distribution that mirrors those of high-resolution images. This evidence underscores the robust capability of our method in reconstructing textures and subtle variations inherent in real-world images, without compromising the retention of intricate details.}

\begin{figure}[H]
\centering
\includegraphics[width=\textwidth]{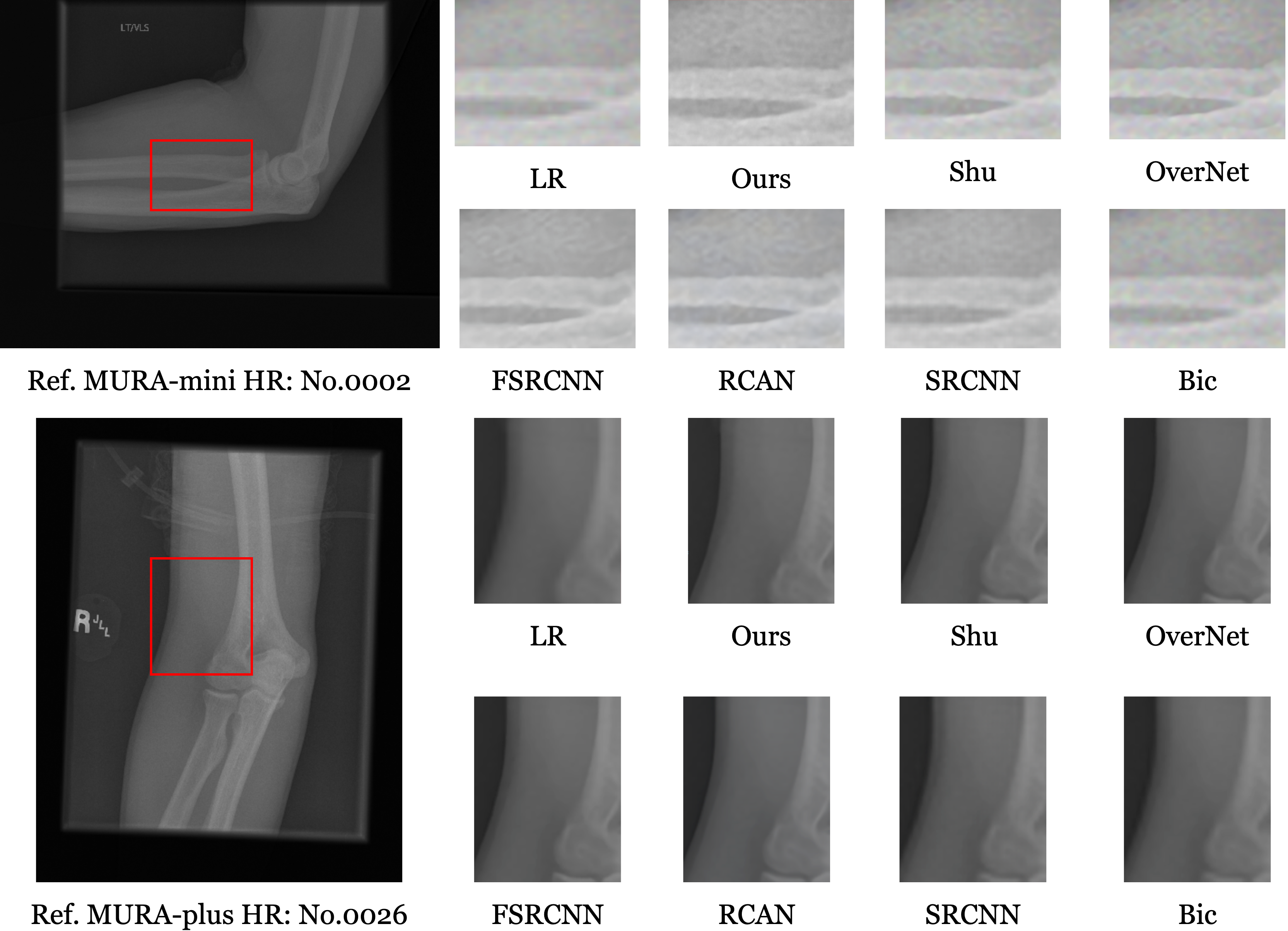}
\caption{Visual comparison achieved on radiology images for \textcolor{red}{$\times 4$} SR.} 
\label{fig6}  
\end{figure}

\begin{figure}[H]
\centering
\includegraphics[width=\textwidth]{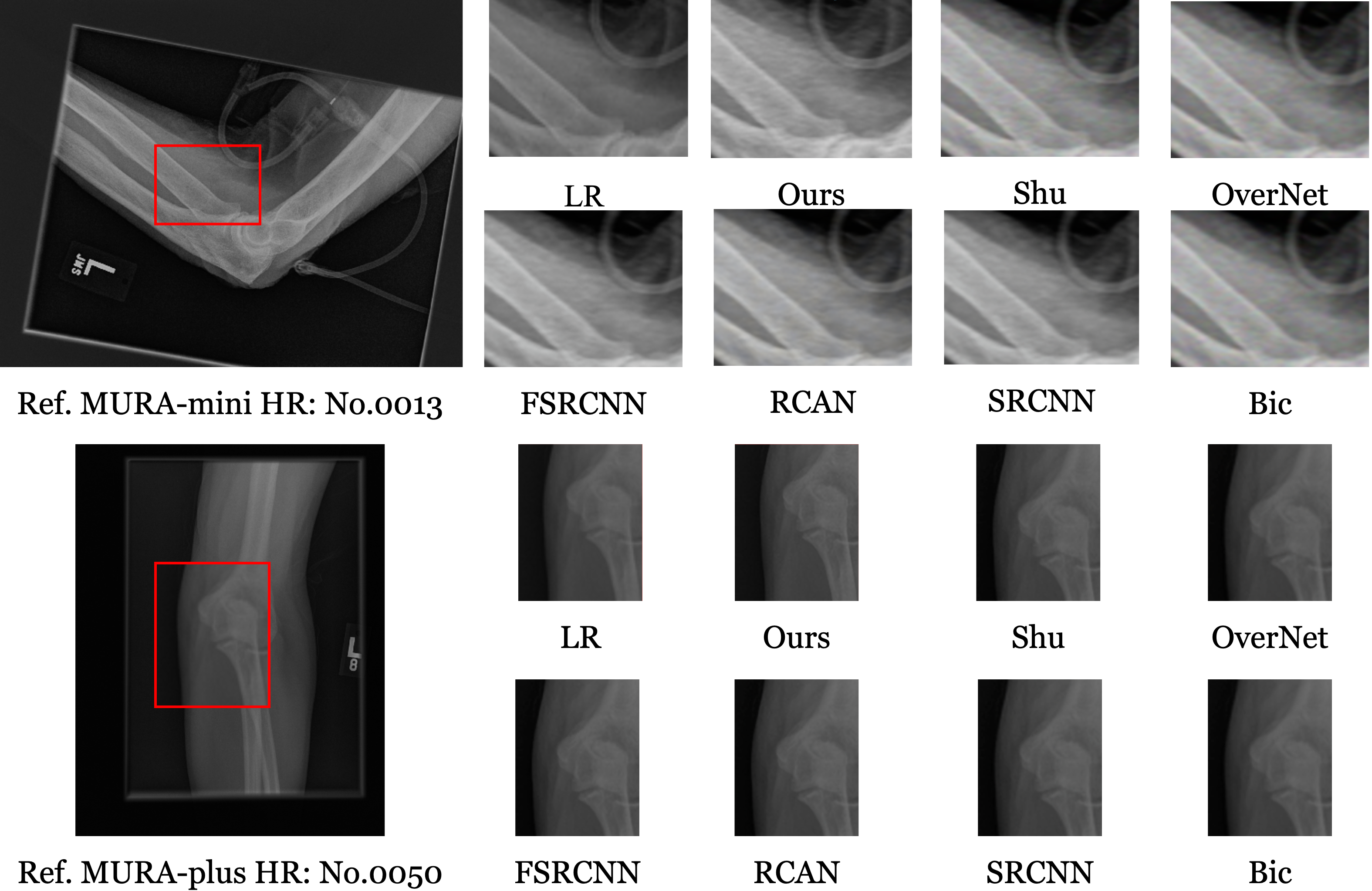}
\caption{Visual comparison achieved on radiology images for \textcolor{red}{$\times 2$} SR.} 
\label{fig7}  
\end{figure}

Secondly, we provide visual comparisons with other SISR methods in Fig. \ref{fig6}-\ref{fig7}. For a more nuanced qualitative analysis, we focus on the highlighted regions. It's crucial to note that the PSNR/SSIM metrics are derived from the entire input image. Observations indicate that our model excels in image restoration. 

\rr{As illustrated in Fig.\ref{fig6}, the $O^{2}$former stands to offer clinicians a more precise insight into aspects such as bone mineral density (Ref. MURA-mini HR: No.0002), bone and joint structure (Ref. MURA-plus HR: No.0026). These enhancements are instrumental for the early identification of changes in bone mineral density (see Fig.\ref{fig7} Ref. MURA-mini HR: No.0013), subtle fractures, and irregularities in joint structures (see Fig.\ref{fig7} Ref. MURA-mini HR: No.0050). Based on the above advantages, our model has a high prospect in clinical application, providing reliable image analysis results for improving early diagnosis and treatment.}



\section{Conclusion\label{sec.5}}
In this study, we proposed a novel approach: $O^{2}$former for radiographic images, with a focus on developing a specific pipeline that can more effectively learn denoising mapping. Given the limited color space and inter-pixel patterns in radiographic images, they cannot be directly compared to normal images. Normal images can more readily learn SR and denoising mapping from inherent patterns, thus radiographic images require enhanced latent representations to aid the decoder's learning process. To achieve this, we introduced a novel orientation operator in the encoder to incorporate the orientation prior and boost the sensitivity to denoising mapping. Additionally, we proposed a multi-scale feature fusion strategy to combine features captured by different receptive fields with the directional prior, thereby providing a superior latent representation for the decoder. Ultimately, we proposed a transformer-based SR model, that is $O^{2}$former, for radiographic images, built upon these two innovative components. Our experimental results demonstrate that our approach outperforms competitors in both qualitative and quantitative comparisons.



\section*{Declaration of Competing Interest}
The authors declare that they have no known competing financial interests or personal relationships that could have appeared to influence the work reported in this paper

\section*{Acknowledgements}
This work was supported by JSPS KAKENHI Grant Number JP23KJ0118.



\bibliographystyle{elsarticle-num-names.bst} 
\bibliography{elbnet}

\begin{thebibliography}{62}
\expandafter\ifx\csname natexlab\endcsname\relax\def\natexlab#1{#1}\fi
\providecommand{\url}[1]{\texttt{#1}}
\providecommand{\href}[2]{#2}
\providecommand{\path}[1]{#1}
\providecommand{\DOIprefix}{doi:}
\providecommand{\ArXivprefix}{arXiv:}
\providecommand{\URLprefix}{URL: }
\providecommand{\Pubmedprefix}{pmid:}
\providecommand{\doi}[1]{\href{http://dx.doi.org/#1}{\path{#1}}}
\providecommand{\Pubmed}[1]{\href{pmid:#1}{\path{#1}}}
\providecommand{\bibinfo}[2]{#2}
\ifx\xfnm\relax \def\xfnm[#1]{\unskip,\space#1}\fi
\bibitem[{Vives(2006)}]{vives2006orthopedic}
\bibinfo{author}{M.~J. Vives},
\newblock \bibinfo{title}{Orthopedic imaging: A practical approach},
\newblock \bibinfo{journal}{The Journal of Spinal Cord Medicine}
  \bibinfo{volume}{29} (\bibinfo{year}{2006}) \bibinfo{pages}{173}.
\bibitem[{Chen et~al.(2013)Chen, Zhou, Fujita, Onozuka, and Kubo}]{chen2013age}
\bibinfo{author}{H.~Chen}, \bibinfo{author}{X.~Zhou},
  \bibinfo{author}{H.~Fujita}, \bibinfo{author}{M.~Onozuka},
  \bibinfo{author}{K.-Y. Kubo},
\newblock \bibinfo{title}{Age-related changes in trabecular and cortical bone
  microstructure},
\newblock \bibinfo{journal}{International journal of endocrinology}
  \bibinfo{volume}{2013} (\bibinfo{year}{2013}).
\bibitem[{Mc~Donnell et~al.(2007)Mc~Donnell, Mc~Hugh, and
  O’mahoney}]{mc2007vertebral}
\bibinfo{author}{P.~Mc~Donnell}, \bibinfo{author}{P.~Mc~Hugh},
  \bibinfo{author}{D.~O’mahoney},
\newblock \bibinfo{title}{Vertebral osteoporosis and trabecular bone quality},
\newblock \bibinfo{journal}{Annals of biomedical engineering}
  \bibinfo{volume}{35} (\bibinfo{year}{2007}) \bibinfo{pages}{170--189}.
\bibitem[{Zhou et~al.(2015)Zhou, Lebel, Treit, Evans, and
  Beaulieu}]{zhou2015accelerated}
\bibinfo{author}{D.~Zhou}, \bibinfo{author}{C.~Lebel},
  \bibinfo{author}{S.~Treit}, \bibinfo{author}{A.~Evans},
  \bibinfo{author}{C.~Beaulieu},
\newblock \bibinfo{title}{Accelerated longitudinal cortical thinning in
  adolescence},
\newblock \bibinfo{journal}{Neuroimage} \bibinfo{volume}{104}
  (\bibinfo{year}{2015}) \bibinfo{pages}{138--145}.
\bibitem[{Turlington(2003)}]{turlington2003radiology}
\bibinfo{author}{B.~Turlington},
\newblock \bibinfo{title}{The radiology of emergency medicine},
\newblock \bibinfo{journal}{Chest} \bibinfo{volume}{123} (\bibinfo{year}{2003})
  \bibinfo{pages}{658}.
\bibitem[{Adepu et~al.(2022)Adepu, Ekman, Leth, Johansson, Lindahl, and
  Ski{\"o}ldebrand}]{adepu2022biglycan}
\bibinfo{author}{S.~Adepu}, \bibinfo{author}{S.~Ekman},
  \bibinfo{author}{J.~Leth}, \bibinfo{author}{U.~Johansson},
  \bibinfo{author}{A.~Lindahl}, \bibinfo{author}{E.~Ski{\"o}ldebrand},
\newblock \bibinfo{title}{Biglycan neo-epitope (bgn262), a novel biomarker for
  screening early changes in equine osteoarthritic subchondral bone},
\newblock \bibinfo{journal}{Osteoarthritis and Cartilage} \bibinfo{volume}{30}
  (\bibinfo{year}{2022}) \bibinfo{pages}{1328--1336}.
\bibitem[{Ying et~al.(2023)Ying, Wang, Shi, Xu, Ge, Sun, Wang, Li, Wu, Tong
  et~al.}]{ying2023inflammation}
\bibinfo{author}{J.~Ying}, \bibinfo{author}{P.~Wang}, \bibinfo{author}{Z.~Shi},
  \bibinfo{author}{J.~Xu}, \bibinfo{author}{Q.~Ge}, \bibinfo{author}{Q.~Sun},
  \bibinfo{author}{W.~Wang}, \bibinfo{author}{J.~Li}, \bibinfo{author}{C.~Wu},
  \bibinfo{author}{P.~Tong}, et~al.,
\newblock \bibinfo{title}{Inflammation-mediated aberrant glucose metabolism in
  subchondral bone induces osteoarthritis},
\newblock \bibinfo{journal}{Stem Cells} \bibinfo{volume}{41}
  (\bibinfo{year}{2023}) \bibinfo{pages}{482--492}.
\bibitem[{Miyamoto et~al.(2007)Miyamoto, Vernotico, Majmundar, and
  Thomas}]{miyamoto2007pharmacologic}
\bibinfo{author}{M.~I. Miyamoto}, \bibinfo{author}{S.~L. Vernotico},
  \bibinfo{author}{H.~Majmundar}, \bibinfo{author}{G.~S. Thomas},
\newblock \bibinfo{title}{Pharmacologic stress myocardial perfusion imaging: a
  practical approach},
\newblock \bibinfo{journal}{Journal of nuclear cardiology} \bibinfo{volume}{14}
  (\bibinfo{year}{2007}) \bibinfo{pages}{250--255}.
\bibitem[{Hu et~al.(2022)Hu, Liu, and Zhang}]{hu2022advance}
\bibinfo{author}{L.~Hu}, \bibinfo{author}{R.~Liu}, \bibinfo{author}{L.~Zhang},
\newblock \bibinfo{title}{Advance in bone destruction participated by jak/stat
  in rheumatoid arthritis and therapeutic effect of jak/stat inhibitors},
\newblock \bibinfo{journal}{International Immunopharmacology}
  \bibinfo{volume}{111} (\bibinfo{year}{2022}) \bibinfo{pages}{109095}.
\bibitem[{Shen and Lv(2022)}]{shen2022dual}
\bibinfo{author}{Y.~Shen}, \bibinfo{author}{Y.~Lv},
\newblock \bibinfo{title}{Dual targeted zeolitic imidazolate framework
  nanoparticles for treating metastatic breast cancer and inhibiting bone
  destruction},
\newblock \bibinfo{journal}{Colloids and Surfaces B: Biointerfaces}
  \bibinfo{volume}{219} (\bibinfo{year}{2022}) \bibinfo{pages}{112826}.
\bibitem[{Shin et~al.(2023)Shin, Peng, Kim, Yoo, and
  Yoon}]{shin2023multivariable}
\bibinfo{author}{M.~Shin}, \bibinfo{author}{Z.~Peng}, \bibinfo{author}{H.-J.
  Kim}, \bibinfo{author}{S.-S. Yoo}, \bibinfo{author}{K.~Yoon},
\newblock \bibinfo{title}{Multivariable-incorporating super-resolution residual
  network for transcranial focused ultrasound simulation},
\newblock \bibinfo{journal}{Computer Methods and Programs in Biomedicine}
  \bibinfo{volume}{237} (\bibinfo{year}{2023}) \bibinfo{pages}{107591}.
\bibitem[{Qiu et~al.(2022)Qiu, Cheng, and Wang}]{qiu2022improved}
\bibinfo{author}{D.~Qiu}, \bibinfo{author}{Y.~Cheng},
  \bibinfo{author}{X.~Wang},
\newblock \bibinfo{title}{Improved generative adversarial network for retinal
  image super-resolution},
\newblock \bibinfo{journal}{Computer Methods and Programs in Biomedicine}
  \bibinfo{volume}{225} (\bibinfo{year}{2022}) \bibinfo{pages}{106995}.
\bibitem[{Zhu et~al.(2023)Zhu, He, and Wang}]{zhu2023feedback}
\bibinfo{author}{D.~Zhu}, \bibinfo{author}{H.~He}, \bibinfo{author}{D.~Wang},
\newblock \bibinfo{title}{Feedback attention network for cardiac magnetic
  resonance imaging super-resolution},
\newblock \bibinfo{journal}{Computer Methods and Programs in Biomedicine}
  \bibinfo{volume}{231} (\bibinfo{year}{2023}) \bibinfo{pages}{107313}.
\bibitem[{Huang et~al.(2023)Huang, Xie, Li, Xiao, You, and
  Liu}]{huang2023source}
\bibinfo{author}{Y.~Huang}, \bibinfo{author}{W.~Xie}, \bibinfo{author}{M.~Li},
  \bibinfo{author}{E.~Xiao}, \bibinfo{author}{J.~You},
  \bibinfo{author}{X.~Liu},
\newblock \bibinfo{title}{Source-free domain adaptive segmentation with
  class-balanced complementary self-training},
\newblock \bibinfo{journal}{Artificial Intelligence in Medicine}
  \bibinfo{volume}{146} (\bibinfo{year}{2023}) \bibinfo{pages}{102694}.
\bibitem[{Wang et~al.(2020)Wang, Chen, and Hoi}]{wang2020deep}
\bibinfo{author}{Z.~Wang}, \bibinfo{author}{J.~Chen}, \bibinfo{author}{S.~C.
  Hoi},
\newblock \bibinfo{title}{Deep learning for image super-resolution: A survey},
\newblock \bibinfo{journal}{IEEE transactions on pattern analysis and machine
  intelligence} \bibinfo{volume}{43} (\bibinfo{year}{2020})
  \bibinfo{pages}{3365--3387}.
\bibitem[{Park et~al.(2003)Park, Park, and Kang}]{park2003super}
\bibinfo{author}{S.~C. Park}, \bibinfo{author}{M.~K. Park},
  \bibinfo{author}{M.~G. Kang},
\newblock \bibinfo{title}{Super-resolution image reconstruction: a technical
  overview},
\newblock \bibinfo{journal}{IEEE signal processing magazine}
  \bibinfo{volume}{20} (\bibinfo{year}{2003}) \bibinfo{pages}{21--36}.
\bibitem[{Chen et~al.(2022)Chen, He, Qing, Wu, Ren, Sheriff, and
  Zhu}]{chen2022real}
\bibinfo{author}{H.~Chen}, \bibinfo{author}{X.~He}, \bibinfo{author}{L.~Qing},
  \bibinfo{author}{Y.~Wu}, \bibinfo{author}{C.~Ren}, \bibinfo{author}{R.~E.
  Sheriff}, \bibinfo{author}{C.~Zhu},
\newblock \bibinfo{title}{Real-world single image super-resolution: A brief
  review},
\newblock \bibinfo{journal}{Information Fusion} \bibinfo{volume}{79}
  (\bibinfo{year}{2022}) \bibinfo{pages}{124--145}.
\bibitem[{Huang et~al.(2022)Huang, Miyazaki, Liu, and
  Omachi}]{huang2022infrared}
\bibinfo{author}{Y.~Huang}, \bibinfo{author}{T.~Miyazaki},
  \bibinfo{author}{X.~Liu}, \bibinfo{author}{S.~Omachi},
\newblock \bibinfo{title}{Infrared image super-resolution: Systematic review,
  and future trends},
\newblock \bibinfo{journal}{arXiv preprint arXiv:2212.12322}
  (\bibinfo{year}{2022}).
\bibitem[{Dong et~al.(2014)Dong, Loy, He, and Tang}]{dong2014learning}
\bibinfo{author}{C.~Dong}, \bibinfo{author}{C.~C. Loy},
  \bibinfo{author}{K.~He}, \bibinfo{author}{X.~Tang},
\newblock \bibinfo{title}{Learning a deep convolutional network for image
  super-resolution},
\newblock in: \bibinfo{booktitle}{Computer Vision--ECCV 2014: 13th European
  Conference, Zurich, Switzerland, September 6-12, 2014, Proceedings, Part IV
  13}, \bibinfo{organization}{Springer}, \bibinfo{year}{2014}, pp.
  \bibinfo{pages}{184--199}.
\bibitem[{Kim et~al.(2016)Kim, Lee, and Lee}]{kim2016accurate}
\bibinfo{author}{J.~Kim}, \bibinfo{author}{J.~K. Lee}, \bibinfo{author}{K.~M.
  Lee},
\newblock \bibinfo{title}{Accurate image super-resolution using very deep
  convolutional networks},
\newblock in: \bibinfo{booktitle}{Proceedings of the IEEE conference on
  computer vision and pattern recognition}, \bibinfo{year}{2016}, pp.
  \bibinfo{pages}{1646--1654}.
\bibitem[{Jiang et~al.(2021)Jiang, Pi, Huang, Qian, and
  Zhang}]{jiang2021difference}
\bibinfo{author}{Z.~Jiang}, \bibinfo{author}{K.~Pi},
  \bibinfo{author}{Y.~Huang}, \bibinfo{author}{Y.~Qian},
  \bibinfo{author}{S.~Zhang},
\newblock \bibinfo{title}{Difference value network for image super-resolution},
\newblock \bibinfo{journal}{IEEE Signal Processing Letters}
  \bibinfo{volume}{28} (\bibinfo{year}{2021}) \bibinfo{pages}{1070--1074}.
\bibitem[{Zhang et~al.(2018)Zhang, Li, Li, Wang, Zhong, and
  Fu}]{zhang2018image}
\bibinfo{author}{Y.~Zhang}, \bibinfo{author}{K.~Li}, \bibinfo{author}{K.~Li},
  \bibinfo{author}{L.~Wang}, \bibinfo{author}{B.~Zhong},
  \bibinfo{author}{Y.~Fu},
\newblock \bibinfo{title}{Image super-resolution using very deep residual
  channel attention networks},
\newblock in: \bibinfo{booktitle}{Proceedings of the European conference on
  computer vision (ECCV)}, \bibinfo{year}{2018}, pp. \bibinfo{pages}{286--301}.
\bibitem[{Behjati et~al.(2021)Behjati, Rodriguez, Mehri, Hupont, Tena, and
  Gonzalez}]{behjati2021overnet}
\bibinfo{author}{P.~Behjati}, \bibinfo{author}{P.~Rodriguez},
  \bibinfo{author}{A.~Mehri}, \bibinfo{author}{I.~Hupont},
  \bibinfo{author}{C.~F. Tena}, \bibinfo{author}{J.~Gonzalez},
\newblock \bibinfo{title}{Overnet: Lightweight multi-scale super-resolution
  with overscaling network},
\newblock in: \bibinfo{booktitle}{Proceedings of the IEEE/CVF Winter Conference
  on Applications of Computer Vision}, \bibinfo{year}{2021}, pp.
  \bibinfo{pages}{2694--2703}.
\bibitem[{Goodfellow et~al.(2020)Goodfellow, Pouget-Abadie, Mirza, Xu,
  Warde-Farley, Ozair, Courville, and Bengio}]{goodfellow2020generative}
\bibinfo{author}{I.~Goodfellow}, \bibinfo{author}{J.~Pouget-Abadie},
  \bibinfo{author}{M.~Mirza}, \bibinfo{author}{B.~Xu},
  \bibinfo{author}{D.~Warde-Farley}, \bibinfo{author}{S.~Ozair},
  \bibinfo{author}{A.~Courville}, \bibinfo{author}{Y.~Bengio},
\newblock \bibinfo{title}{Generative adversarial networks},
\newblock \bibinfo{journal}{Communications of the ACM} \bibinfo{volume}{63}
  (\bibinfo{year}{2020}) \bibinfo{pages}{139--144}.
\bibitem[{Ledig et~al.(2017)Ledig, Theis, Husz{\'a}r, Caballero, Cunningham,
  Acosta, Aitken, Tejani, Totz, Wang et~al.}]{ledig2017photo}
\bibinfo{author}{C.~Ledig}, \bibinfo{author}{L.~Theis},
  \bibinfo{author}{F.~Husz{\'a}r}, \bibinfo{author}{J.~Caballero},
  \bibinfo{author}{A.~Cunningham}, \bibinfo{author}{A.~Acosta},
  \bibinfo{author}{A.~Aitken}, \bibinfo{author}{A.~Tejani},
  \bibinfo{author}{J.~Totz}, \bibinfo{author}{Z.~Wang}, et~al.,
\newblock \bibinfo{title}{Photo-realistic single image super-resolution using a
  generative adversarial network},
\newblock in: \bibinfo{booktitle}{Proceedings of the IEEE conference on
  computer vision and pattern recognition}, \bibinfo{year}{2017}, pp.
  \bibinfo{pages}{4681--4690}.
\bibitem[{Huang et~al.(2021)Huang, Jiang, Lan, Zhang, and
  Pi}]{huang2021infrared}
\bibinfo{author}{Y.~Huang}, \bibinfo{author}{Z.~Jiang},
  \bibinfo{author}{R.~Lan}, \bibinfo{author}{S.~Zhang},
  \bibinfo{author}{K.~Pi},
\newblock \bibinfo{title}{Infrared image super-resolution via transfer learning
  and psrgan},
\newblock \bibinfo{journal}{IEEE Signal Processing Letters}
  \bibinfo{volume}{28} (\bibinfo{year}{2021}) \bibinfo{pages}{982--986}.
\bibitem[{Wang et~al.(2018)Wang, Yu, Wu, Gu, Liu, Dong, Qiao, and
  Change~Loy}]{wang2018esrgan}
\bibinfo{author}{X.~Wang}, \bibinfo{author}{K.~Yu}, \bibinfo{author}{S.~Wu},
  \bibinfo{author}{J.~Gu}, \bibinfo{author}{Y.~Liu}, \bibinfo{author}{C.~Dong},
  \bibinfo{author}{Y.~Qiao}, \bibinfo{author}{C.~Change~Loy},
\newblock \bibinfo{title}{Esrgan: Enhanced super-resolution generative
  adversarial networks},
\newblock in: \bibinfo{booktitle}{Proceedings of the European conference on
  computer vision (ECCV) workshops}, \bibinfo{year}{2018}, pp.
  \bibinfo{pages}{0--0}.
\bibitem[{Wang et~al.(2021)Wang, Xie, Dong, and Shan}]{wang2021real}
\bibinfo{author}{X.~Wang}, \bibinfo{author}{L.~Xie}, \bibinfo{author}{C.~Dong},
  \bibinfo{author}{Y.~Shan},
\newblock \bibinfo{title}{Real-esrgan: Training real-world blind
  super-resolution with pure synthetic data},
\newblock in: \bibinfo{booktitle}{Proceedings of the IEEE/CVF International
  Conference on Computer Vision}, \bibinfo{year}{2021}, pp.
  \bibinfo{pages}{1905--1914}.
\bibitem[{Gulrajani et~al.(2017)Gulrajani, Ahmed, Arjovsky, Dumoulin, and
  Courville}]{gulrajani2017improved}
\bibinfo{author}{I.~Gulrajani}, \bibinfo{author}{F.~Ahmed},
  \bibinfo{author}{M.~Arjovsky}, \bibinfo{author}{V.~Dumoulin},
  \bibinfo{author}{A.~C. Courville},
\newblock \bibinfo{title}{Improved training of wasserstein gans},
\newblock \bibinfo{journal}{Advances in neural information processing systems}
  \bibinfo{volume}{30} (\bibinfo{year}{2017}).
\bibitem[{Yang et~al.(2020)Yang, Yang, Fu, Lu, and Guo}]{yang2020learning}
\bibinfo{author}{F.~Yang}, \bibinfo{author}{H.~Yang}, \bibinfo{author}{J.~Fu},
  \bibinfo{author}{H.~Lu}, \bibinfo{author}{B.~Guo},
\newblock \bibinfo{title}{Learning texture transformer network for image
  super-resolution},
\newblock in: \bibinfo{booktitle}{Proceedings of the IEEE/CVF conference on
  computer vision and pattern recognition}, \bibinfo{year}{2020}, pp.
  \bibinfo{pages}{5791--5800}.
\bibitem[{Lu et~al.(2022)Lu, Li, Liu, Huang, Zhang, and
  Zeng}]{lu2022transformer}
\bibinfo{author}{Z.~Lu}, \bibinfo{author}{J.~Li}, \bibinfo{author}{H.~Liu},
  \bibinfo{author}{C.~Huang}, \bibinfo{author}{L.~Zhang},
  \bibinfo{author}{T.~Zeng},
\newblock \bibinfo{title}{Transformer for single image super-resolution},
\newblock in: \bibinfo{booktitle}{Proceedings of the IEEE/CVF Conference on
  Computer Vision and Pattern Recognition}, \bibinfo{year}{2022}, pp.
  \bibinfo{pages}{457--466}.
\bibitem[{Gao et~al.(2022)Gao, Wang, Li, Li, Yu, and Zeng}]{gao2022lightweight}
\bibinfo{author}{G.~Gao}, \bibinfo{author}{Z.~Wang}, \bibinfo{author}{J.~Li},
  \bibinfo{author}{W.~Li}, \bibinfo{author}{Y.~Yu}, \bibinfo{author}{T.~Zeng},
\newblock \bibinfo{title}{Lightweight bimodal network for single-image
  super-resolution via symmetric cnn and recursive transformer},
\newblock \bibinfo{journal}{arXiv preprint arXiv:2204.13286}
  (\bibinfo{year}{2022}).
\bibitem[{Qiu et~al.(2023)Qiu, Cheng, and Wang}]{qiu2023medical}
\bibinfo{author}{D.~Qiu}, \bibinfo{author}{Y.~Cheng},
  \bibinfo{author}{X.~Wang},
\newblock \bibinfo{title}{Medical image super-resolution reconstruction
  algorithms based on deep learning: A survey},
\newblock \bibinfo{journal}{Computer Methods and Programs in Biomedicine}
  (\bibinfo{year}{2023}) \bibinfo{pages}{107590}.
\bibitem[{Qiu et~al.(2022)Qiu, Cheng, and Wang}]{qiu2022dual}
\bibinfo{author}{D.~Qiu}, \bibinfo{author}{Y.~Cheng},
  \bibinfo{author}{X.~Wang},
\newblock \bibinfo{title}{Dual u-net residual networks for cardiac magnetic
  resonance images super-resolution},
\newblock \bibinfo{journal}{Computer Methods and Programs in Biomedicine}
  \bibinfo{volume}{218} (\bibinfo{year}{2022}) \bibinfo{pages}{106707}.
\bibitem[{Zhu and Qiu(2021)}]{zhu2021residual}
\bibinfo{author}{D.~Zhu}, \bibinfo{author}{D.~Qiu},
\newblock \bibinfo{title}{Residual dense network for medical magnetic resonance
  images super-resolution},
\newblock \bibinfo{journal}{Computer Methods and Programs in Biomedicine}
  \bibinfo{volume}{209} (\bibinfo{year}{2021}) \bibinfo{pages}{106330}.
\bibitem[{Dong et~al.(2015)Dong, Loy, He, and Tang}]{dong2015image}
\bibinfo{author}{C.~Dong}, \bibinfo{author}{C.~C. Loy},
  \bibinfo{author}{K.~He}, \bibinfo{author}{X.~Tang},
\newblock \bibinfo{title}{Image super-resolution using deep convolutional
  networks},
\newblock \bibinfo{journal}{IEEE transactions on pattern analysis and machine
  intelligence} \bibinfo{volume}{38} (\bibinfo{year}{2015})
  \bibinfo{pages}{295--307}.
\bibitem[{Huang et~al.(2021)Huang, Jiang, Wang, Jiang, and Pang}]{huang2021het}
\bibinfo{author}{Y.~Huang}, \bibinfo{author}{Z.~Jiang},
  \bibinfo{author}{Q.~Wang}, \bibinfo{author}{Q.~Jiang},
  \bibinfo{author}{G.~Pang},
\newblock \bibinfo{title}{Infrared image super-resolution via heterogeneous
  convolutional wgan},
\newblock in: \bibinfo{booktitle}{PRICAI 2021: Trends in Artificial
  Intelligence: 18th Pacific Rim International Conference on Artificial
  Intelligence, PRICAI 2021, Hanoi, Vietnam, November 8--12, 2021, Proceedings,
  Part II 18}, \bibinfo{organization}{Springer}, \bibinfo{year}{2021}, pp.
  \bibinfo{pages}{461--472}.
\bibitem[{Huang et~al.(2023)Huang, Miyazaki, Liu, Dong, and
  Omachi}]{huang2023target}
\bibinfo{author}{Y.~Huang}, \bibinfo{author}{T.~Miyazaki},
  \bibinfo{author}{X.~Liu}, \bibinfo{author}{Y.~Dong},
  \bibinfo{author}{S.~Omachi},
\newblock \bibinfo{title}{Target-oriented domain adaptation for infrared image
  super-resolution},
\newblock \bibinfo{journal}{arXiv preprint arXiv:2311.08816}
  (\bibinfo{year}{2023}).
\bibitem[{Han et~al.(2020)Han, Wang, Chen, Chen, Guo, Liu, Tang, Xiao, Xu, Xu
  et~al.}]{han2020survey}
\bibinfo{author}{K.~Han}, \bibinfo{author}{Y.~Wang}, \bibinfo{author}{H.~Chen},
  \bibinfo{author}{X.~Chen}, \bibinfo{author}{J.~Guo},
  \bibinfo{author}{Z.~Liu}, \bibinfo{author}{Y.~Tang},
  \bibinfo{author}{A.~Xiao}, \bibinfo{author}{C.~Xu}, \bibinfo{author}{Y.~Xu},
  et~al.,
\newblock \bibinfo{title}{A survey on visual transformer},
\newblock \bibinfo{journal}{arXiv preprint arXiv:2012.12556}
  \bibinfo{volume}{2} (\bibinfo{year}{2020}).
\bibitem[{Vaswani et~al.(2017)Vaswani, Shazeer, Parmar, Uszkoreit, Jones,
  Gomez, Kaiser, and Polosukhin}]{vaswani2017attention}
\bibinfo{author}{A.~Vaswani}, \bibinfo{author}{N.~Shazeer},
  \bibinfo{author}{N.~Parmar}, \bibinfo{author}{J.~Uszkoreit},
  \bibinfo{author}{L.~Jones}, \bibinfo{author}{A.~N. Gomez},
  \bibinfo{author}{{\L}.~Kaiser}, \bibinfo{author}{I.~Polosukhin},
\newblock \bibinfo{title}{Attention is all you need},
\newblock \bibinfo{journal}{Advances in neural information processing systems}
  \bibinfo{volume}{30} (\bibinfo{year}{2017}).
\bibitem[{He et~al.(2023)He, Chen, Cao, Yang, Cao, Li, Tang, Zhuang, and
  Lu}]{he2023single}
\bibinfo{author}{Z.~He}, \bibinfo{author}{D.~Chen}, \bibinfo{author}{Y.~Cao},
  \bibinfo{author}{J.~Yang}, \bibinfo{author}{Y.~Cao}, \bibinfo{author}{X.~Li},
  \bibinfo{author}{S.~Tang}, \bibinfo{author}{Y.~Zhuang},
  \bibinfo{author}{Z.-m. Lu},
\newblock \bibinfo{title}{Single image super-resolution based on progressive
  fusion of orientation-aware features},
\newblock \bibinfo{journal}{Pattern Recognition} \bibinfo{volume}{133}
  (\bibinfo{year}{2023}) \bibinfo{pages}{109038}.
\bibitem[{Lin et~al.(2023)Lin, Zhang, Fang, Chen, Cheng, and
  Chen}]{lin2023rethinking}
\bibinfo{author}{Y.~Lin}, \bibinfo{author}{D.~Zhang},
  \bibinfo{author}{X.~Fang}, \bibinfo{author}{Y.~Chen}, \bibinfo{author}{K.-T.
  Cheng}, \bibinfo{author}{H.~Chen},
\newblock \bibinfo{title}{Rethinking boundary detection in deep learning models
  for medical image segmentation},
\newblock in: \bibinfo{booktitle}{International Conference on Information
  Processing in Medical Imaging}, \bibinfo{organization}{Springer},
  \bibinfo{year}{2023}, pp. \bibinfo{pages}{730--742}.
\bibitem[{Vo and Kim(2023)}]{vo2023mulvernet}
\bibinfo{author}{V.~T.-T. Vo}, \bibinfo{author}{S.-H. Kim},
\newblock \bibinfo{title}{Mulvernet: Nucleus segmentation and classification of
  pathology images using the hover-net and multiple filter units},
\newblock \bibinfo{journal}{Electronics} \bibinfo{volume}{12}
  (\bibinfo{year}{2023}) \bibinfo{pages}{355}.
\bibitem[{Dogar et~al.(2023)Dogar, Shahzad, and Fraz}]{dogar2023attention}
\bibinfo{author}{G.~M. Dogar}, \bibinfo{author}{M.~Shahzad},
  \bibinfo{author}{M.~M. Fraz},
\newblock \bibinfo{title}{Attention augmented distance regression and
  classification network for nuclei instance segmentation and type
  classification in histology images},
\newblock \bibinfo{journal}{Biomedical Signal Processing and Control}
  \bibinfo{volume}{79} (\bibinfo{year}{2023}) \bibinfo{pages}{104199}.
\bibitem[{Tsai et~al.(2022)Tsai, Peng, Lin, Tsai, and
  Lin}]{tsai2022stripformer}
\bibinfo{author}{F.-J. Tsai}, \bibinfo{author}{Y.-T. Peng},
  \bibinfo{author}{Y.-Y. Lin}, \bibinfo{author}{C.-C. Tsai},
  \bibinfo{author}{C.-W. Lin},
\newblock \bibinfo{title}{Stripformer: Strip transformer for fast image
  deblurring},
\newblock in: \bibinfo{booktitle}{Computer Vision--ECCV 2022: 17th European
  Conference, Tel Aviv, Israel, October 23--27, 2022, Proceedings, Part XIX},
  \bibinfo{organization}{Springer}, \bibinfo{year}{2022}, pp.
  \bibinfo{pages}{146--162}.
\bibitem[{Sun et~al.(2015)Sun, Cao, Xu, and Ponce}]{sun2015learning}
\bibinfo{author}{J.~Sun}, \bibinfo{author}{W.~Cao}, \bibinfo{author}{Z.~Xu},
  \bibinfo{author}{J.~Ponce},
\newblock \bibinfo{title}{Learning a convolutional neural network for
  non-uniform motion blur removal},
\newblock in: \bibinfo{booktitle}{Proceedings of the IEEE conference on
  computer vision and pattern recognition}, \bibinfo{year}{2015}, pp.
  \bibinfo{pages}{769--777}.
\bibitem[{Huang et~al.(2023)Huang, Xie, Li, Cheng, Wu, Wang, You, and
  Liu}]{huang2023vicinal}
\bibinfo{author}{Y.~Huang}, \bibinfo{author}{W.~Xie}, \bibinfo{author}{M.~Li},
  \bibinfo{author}{M.~Cheng}, \bibinfo{author}{J.~Wu},
  \bibinfo{author}{W.~Wang}, \bibinfo{author}{J.~You},
  \bibinfo{author}{X.~Liu},
\newblock \bibinfo{title}{Vicinal feature statistics augmentation for federated
  3d medical volume segmentation},
\newblock in: \bibinfo{booktitle}{International Conference on Information
  Processing in Medical Imaging}, \bibinfo{organization}{Springer},
  \bibinfo{year}{2023}, pp. \bibinfo{pages}{360--371}.
\bibitem[{Zhang et~al.(2018)Zhang, Tian, Kong, Zhong, and
  Fu}]{zhang2018residual}
\bibinfo{author}{Y.~Zhang}, \bibinfo{author}{Y.~Tian},
  \bibinfo{author}{Y.~Kong}, \bibinfo{author}{B.~Zhong},
  \bibinfo{author}{Y.~Fu},
\newblock \bibinfo{title}{Residual dense network for image super-resolution},
\newblock in: \bibinfo{booktitle}{Proceedings of the IEEE conference on
  computer vision and pattern recognition}, \bibinfo{year}{2018}, pp.
  \bibinfo{pages}{2472--2481}.
\bibitem[{Li et~al.(2018)Li, Fang, Mei, and Zhang}]{li2018multi}
\bibinfo{author}{J.~Li}, \bibinfo{author}{F.~Fang}, \bibinfo{author}{K.~Mei},
  \bibinfo{author}{G.~Zhang},
\newblock \bibinfo{title}{Multi-scale residual network for image
  super-resolution},
\newblock in: \bibinfo{booktitle}{Proceedings of the European conference on
  computer vision (ECCV)}, \bibinfo{year}{2018}, pp. \bibinfo{pages}{517--532}.
\bibitem[{Tong et~al.(2017)Tong, Li, Liu, and Gao}]{tong2017image}
\bibinfo{author}{T.~Tong}, \bibinfo{author}{G.~Li}, \bibinfo{author}{X.~Liu},
  \bibinfo{author}{Q.~Gao},
\newblock \bibinfo{title}{Image super-resolution using dense skip connections},
\newblock in: \bibinfo{booktitle}{Proceedings of the IEEE international
  conference on computer vision}, \bibinfo{year}{2017}, pp.
  \bibinfo{pages}{4799--4807}.
\bibitem[{Zhang et~al.(2022)Zhang, Zeng, Guo, and Zhang}]{zhang2022efficient}
\bibinfo{author}{X.~Zhang}, \bibinfo{author}{H.~Zeng},
  \bibinfo{author}{S.~Guo}, \bibinfo{author}{L.~Zhang},
\newblock \bibinfo{title}{Efficient long-range attention network for image
  super-resolution},
\newblock in: \bibinfo{booktitle}{Computer Vision--ECCV 2022: 17th European
  Conference, Tel Aviv, Israel, October 23--27, 2022, Proceedings, Part XVII},
  \bibinfo{organization}{Springer}, \bibinfo{year}{2022}, pp.
  \bibinfo{pages}{649--667}.
\bibitem[{Huang et~al.(2022)Huang, Wang, and Omachi}]{huang2022rethinking}
\bibinfo{author}{Y.~Huang}, \bibinfo{author}{Q.~Wang},
  \bibinfo{author}{S.~Omachi},
\newblock \bibinfo{title}{Rethinking degradation: Radiograph super-resolution
  via aid-srgan},
\newblock in: \bibinfo{booktitle}{Machine Learning in Medical Imaging: 13th
  International Workshop, MLMI 2022, Held in Conjunction with MICCAI 2022,
  Singapore, September 18, 2022, Proceedings},
  \bibinfo{organization}{Springer}, \bibinfo{year}{2022}, pp.
  \bibinfo{pages}{43--52}.
\bibitem[{Kingma and Ba(2014)}]{kingma2014adam}
\bibinfo{author}{D.~P. Kingma}, \bibinfo{author}{J.~Ba},
\newblock \bibinfo{title}{Adam: A method for stochastic optimization},
\newblock \bibinfo{journal}{arXiv preprint arXiv:1412.6980}
  (\bibinfo{year}{2014}).
\bibitem[{Wang et~al.(2004)Wang, Bovik, Sheikh, and Simoncelli}]{wang2004image}
\bibinfo{author}{Z.~Wang}, \bibinfo{author}{A.~C. Bovik},
  \bibinfo{author}{H.~R. Sheikh}, \bibinfo{author}{E.~P. Simoncelli},
\newblock \bibinfo{title}{Image quality assessment: from error visibility to
  structural similarity},
\newblock \bibinfo{journal}{IEEE transactions on image processing}
  \bibinfo{volume}{13} (\bibinfo{year}{2004}) \bibinfo{pages}{600--612}.
\bibitem[{Blau et~al.(2018)Blau, Mechrez, Timofte, Michaeli, and
  Zelnik-Manor}]{blau20182018}
\bibinfo{author}{Y.~Blau}, \bibinfo{author}{R.~Mechrez},
  \bibinfo{author}{R.~Timofte}, \bibinfo{author}{T.~Michaeli},
  \bibinfo{author}{L.~Zelnik-Manor},
\newblock \bibinfo{title}{The 2018 pirm challenge on perceptual image
  super-resolution},
\newblock in: \bibinfo{booktitle}{Proceedings of the European Conference on
  Computer Vision (ECCV) Workshops}, \bibinfo{year}{2018}, pp.
  \bibinfo{pages}{0--0}.
\bibitem[{Mittal et~al.(2012)Mittal, Soundararajan, and
  Bovik}]{mittal2012making}
\bibinfo{author}{A.~Mittal}, \bibinfo{author}{R.~Soundararajan},
  \bibinfo{author}{A.~C. Bovik},
\newblock \bibinfo{title}{Making a “completely blind” image quality
  analyzer},
\newblock \bibinfo{journal}{IEEE Signal processing letters}
  \bibinfo{volume}{20} (\bibinfo{year}{2012}) \bibinfo{pages}{209--212}.
\bibitem[{Behjati et~al.(2023)Behjati, Rodriguez, Fern{\'a}ndez, Hupont, Mehri,
  and Gonz{\`a}lez}]{behjati2023single}
\bibinfo{author}{P.~Behjati}, \bibinfo{author}{P.~Rodriguez},
  \bibinfo{author}{C.~Fern{\'a}ndez}, \bibinfo{author}{I.~Hupont},
  \bibinfo{author}{A.~Mehri}, \bibinfo{author}{J.~Gonz{\`a}lez},
\newblock \bibinfo{title}{Single image super-resolution based on directional
  variance attention network},
\newblock \bibinfo{journal}{Pattern Recognition} \bibinfo{volume}{133}
  (\bibinfo{year}{2023}) \bibinfo{pages}{108997}.
\bibitem[{Sun et~al.(2022)Sun, Pan, and Tang}]{sun2022shufflemixer}
\bibinfo{author}{L.~Sun}, \bibinfo{author}{J.~Pan}, \bibinfo{author}{J.~Tang},
\newblock \bibinfo{title}{Shufflemixer: An efficient convnet for image
  super-resolution},
\newblock \bibinfo{journal}{arXiv preprint arXiv:2205.15175}
  (\bibinfo{year}{2022}).
\bibitem[{Sun et~al.(2023)Sun, Dong, Tang, and Pan}]{sun2023spatially}
\bibinfo{author}{L.~Sun}, \bibinfo{author}{J.~Dong}, \bibinfo{author}{J.~Tang},
  \bibinfo{author}{J.~Pan},
\newblock \bibinfo{title}{Spatially-adaptive feature modulation for efficient
  image super-resolution},
\newblock \bibinfo{journal}{arXiv preprint arXiv:2302.13800}
  (\bibinfo{year}{2023}).
\bibitem[{Dong et~al.(2016)Dong, Loy, and Tang}]{dong2016accelerating}
\bibinfo{author}{C.~Dong}, \bibinfo{author}{C.~C. Loy},
  \bibinfo{author}{X.~Tang},
\newblock \bibinfo{title}{Accelerating the super-resolution convolutional
  neural network},
\newblock in: \bibinfo{booktitle}{Computer Vision--ECCV 2016: 14th European
  Conference, Amsterdam, The Netherlands, October 11-14, 2016, Proceedings,
  Part II 14}, \bibinfo{organization}{Springer}, \bibinfo{year}{2016}, pp.
  \bibinfo{pages}{391--407}.
\bibitem[{Dalal and Triggs(2005)}]{dalal2005histograms}
\bibinfo{author}{N.~Dalal}, \bibinfo{author}{B.~Triggs},
\newblock \bibinfo{title}{Histograms of oriented gradients for human
  detection},
\newblock in: \bibinfo{booktitle}{2005 IEEE computer society conference on
  computer vision and pattern recognition (CVPR'05)},
  volume~\bibinfo{volume}{1}, \bibinfo{organization}{Ieee},
  \bibinfo{year}{2005}, pp. \bibinfo{pages}{886--893}.
\bibitem[{Huang et~al.(2021)Huang, Zhang, Zhang, and Shan}]{huang2021gan}
\bibinfo{author}{Z.~Huang}, \bibinfo{author}{J.~Zhang},
  \bibinfo{author}{Y.~Zhang}, \bibinfo{author}{H.~Shan},
\newblock \bibinfo{title}{Du-gan: Generative adversarial networks with
  dual-domain u-net-based discriminators for low-dose ct denoising},
\newblock \bibinfo{journal}{IEEE Transactions on Instrumentation and
  Measurement} \bibinfo{volume}{71} (\bibinfo{year}{2021})
  \bibinfo{pages}{1--12}.

\end{thebibliography}

\end{document}